\documentclass[12pt,preprint]{aastex}
 
\shorttitle{Massive Star Formation}
\shortauthors{Cunningham et al.}

\def\solar{\ifmmode _{\mathord\odot}\else $_{\mathord\odot}$\fi}
\def\msun{\ifmmode {\rm M}_{\mathord\odot}\else $M_{\mathord\odot}$\fi}
\def\lsun{\ifmmode {\rm L}_{\mathord\odot}\else $L_{\mathord\odot}$\fi}
\def\rsun{\ifmmode {\rm R}_{\mathord\odot}\else $L_{\mathord\odot}$\fi}
\def\sol{\ifmmode {\mathord\odot}\else ${\mathord\odot}$\fi}
\newcommand{\textsubscript}[1]{\textrm{\scriptsize #1}}

\newcommand{\grad}{\nabla}
\newcommand{\atan}{\textnormal{atan}}
\renewcommand{\vec}[1]{{\bf #1}}

\begin{document}
\title{Radiation-Hydrodynamic Simulations of Massive Star Formation with Protostellar Outflows}

\author{Andrew J. Cunningham\altaffilmark{1}, Richard I. Klein\altaffilmark{1,2}, Mark R. Krumholz\altaffilmark{3}, Christopher F. McKee\altaffilmark{2,4}}
\altaffiltext{1}{Lawrence Livermore National Laboratory, Livermore, CA 94550}
\altaffiltext{2}{Department of Astronomy, University of California Berkeley, Berkeley, CA 94720}
\altaffiltext{3}{Department of Astronomy and Astrophysics, University of California Santa Cruz, Santa Cruz, CA 94560}
\altaffiltext{4}{Department of Physics, University of California Berkeley, Berkeley, CA 94720}

\email{ajcunn@gmail.com}

\begin{abstract}
We report the results of a series of AMR radiation-hydrodynamic
simulations of the collapse of massive star forming clouds using the
ORION code. These simulations are the first to include the feedback
effects protostellar outflows, as well as protostellar radiative
heating and radiation pressure exerted on the infalling, dusty gas. We
find that outflows evacuate polar cavities of reduced optical depth
through the ambient core. These enhance the radiative flux in the
poleward direction so that it is 1.7 to 15 times larger than that in
the midplane. As a result the radiative heating and outward radiation
force exerted on the protostellar disk and infalling cloud gas in the
equatorial direction are greatly diminished. This simultaneously
reduces the Eddington radiation pressure barrier to high-mass star
formation and increases the minimum threshold surface density for
radiative heating to suppress fragmentation compared to models that do
not include outflows. The strength of both these effects depends on
the initial core surface density. Lower surface density cores have
longer free-fall times and thus massive stars formed within them
undergo more Kelvin contraction as the core collapses, leading to more
powerful outflows.  Furthermore, in lower surface density clouds the
ratio of the time required for the outflow to break out of the core to
the core free-fall time is smaller, so that these clouds are
consequently influenced by outflows at earlier stages of collapse. As
a result, outflow effects are strongest in low surface density cores
and weakest in high surface density one.  We also find that radiation
focusing in the direction of outflow cavities is sufficient to prevent
the formation of radiation pressure-supported circumstellar gas
bubbles, in contrast to models which neglect protostellar outflow
feedback.
\end{abstract}


\section{Introduction} \label{introduction}
Stars of all masses undergo energetic, bipolar mass loss during their
formation \citep{Shepherd,Richer}.  These outflows feed energy back
into large-scale turbulent motions that support clouds against
collapse, may play a role in dispersing some localized regions entirely
\citep{NormanSilk,McKee89,Nakamura,Carroll,Arce} and regulate the
final mass of the central star \citep{MatznerCores,ArceSargent,Wang}.
Massive stars likely provide the dominant source of radiation feedback
in the evolution of their parent molecular clouds and any subsequent
star formation therein. Although much progress has been made both
observationally and theoretically, a comprehensive picture of massive
star formation and the role of feedback from massive stars in
mediating the star formation process remains to be elucidated
\citep{Zinnecker}.

Observational evidence supports a picture where accretion and outflow
ejection processes at work in the formation of high-mass stars proceed
as a ``scaled-up'' version of their low-mass, solar-type counterparts.
Interferometric molecular line measurements have detected quiescent
compact cores within dense, infrared dark clouds of mass $\sim 100
\msun$ \citep{Swift} as likely candidates to the onset of high-mass
star formation.  Observational surveys have established a correlation
between molecular outflow mass-loss and source luminosity
\citep{ShepherdChurchwell,Richer} and between circumstellar mass and
luminosity from $0.1$ to $10^5\lsun$ \citep{Saraceno,Chandler}.
Several authors have detected molecular outflows from massive
protostars with collimation factors of $2$ to $10$
\citep{Zhang,Beuther02a,Beuther02c,Beuther03,Beuther04,Qui,lopez11},
similar to that of low-mass stars \citep{Bachiller}.  Radio thermal
continuum emission jets, commonly associated with low-mass protostars
\citep{Rodriguez}, have also been identified near protostellar sources
as luminous as $\sim 10^5\lsun$ \citep{Torrelles,Curiel}.  Detection
of synchrotron emission arising from the jet in one massive young
stellar object gives support to the notion of a common magnetic
driving mechanism to protostellar outflow from stars of all masses
\citep{massivemhdjet}.  Because high-mass accretion disks are deeply
embedded in dusty envelopes, they are particularly difficult to
observe directly.  A few such detections have, however, been made by
maser emission sources \citep{Hutawarakorn}, in high-resolution
submillimeter dust emission \citep{Patel} and in near-infrared
observations where winds from nearby sources have cleared dust from
the line of sight \citep{Nurnberger}.  These observations suggest that
massive stars form through disk accretion in direct analogy to the
formation of low-mass stars.

Several key theoretical aspects distinguish high and low-mass star
formation despite the broad similarity of the observed outflow and
ejection phenomena.  Massive stars are shorter-lived and produce more
sources of energetic feedback into their environment than their
low-mass counterparts.  O stars radiate their gravitational binding
energy and reach the main sequence on Kelvin-Helmholtz timescales of
$\lesssim 10^4~\textnormal{yr}$ whereas solar type stars require
$\gtrsim 10^7~\textnormal{yr}$.  Stars with masses $\gtrsim 10\msun$
therefore begin nuclear burning while they are still embedded within
and accreting from the circumstellar envelope \citep{shu87}.  The
resultant spherically-averaged radiation pressure on dust grains in
the infalling gas exceeds the gravitational pull from the central star
\citep{larson}.  Massive stars can therefore only form by accretion if
some mechanism is in place to focus the outward radiative flux away
from the infalling envelope.  A variety of focusing mechanism have
been suggested
\citep{Nakano,Jijina,Yorke,KrumholzRTBubbles,McKeeOstriker,Kuiper},
including the one of central interest for this paper, beaming of
radiation in the cavities produced by protostellar outflows.
\citep{KrumholzOutflows}.  Once the embedded stars reach the main
sequence, ionizing photons generate HII regions, strongly affecting
the physical structure and chemistry of their environment.  Recent
observation suggests the existence of stars as massive as $300\msun$
\citep{Crowther}, and it remains unclear if such large mass can be
reached by accretion alone in spite of these strong feedback effects.
Massive stars appear predominantly in denser clusters than low-mass
stars, and massive stars more frequently occur in binary and
small-multiple systems \citep{Preibisch,Shatsky,Lada}.  Furthermore,
recent theoretical \citep{Krumholz08}, numerical \citep{krumholz10}
and observational \citep{lopez} evidence indicate a minimum prestellar
core surface density for high-mass star formation, giving rise to a
specific environmental dependence that distinguishes the case of
massive star formation.

In this paper we present a series of AMR radiation-hydrodynamic
simulations of the collapse of massive star forming clouds using the
ORION code \citep{Truelovethesis,Truelove,Klein,KrumholzOrion}.  These
simulations are the first to simultaneously include radiation and
protostellar outflow feedback, and to study their interaction.  This
work is complementary to that of \cite{Peters}, which included the
effect of photoionization but not of radiation pressure or outflows.
To probe the environmental dependence for massive star formation, we
examine the effect of outflows in star forming cores at several
surface densities representative of typical massive star forming
regions in the Milky Way to regions characteristic of extragalactic
super star clusters.  To isolate the effect of outflow feedback alone,
we include one model where outflows have been turned off in an
otherwise identical cloud. In \S\ref{setup} we describe the simulation
methodology and input parameters, in \S\ref{results} the numerical
results are presented and discussed and in \S\ref{summary} we
summarize the conclusions that can be drawn from the models.

\section{Simulation Setup} \label{setup}
\subsection{Initial Conditions}
Our simulations are initialized to the state of a prestellar core of
mass M, each with a power law density profile given by
\begin{equation}
\rho(r) \propto r^{-k\rho},
\end{equation}
with $k_\rho=3/2$, consistent with models \citep{McKeeTan02,McKeeTan}
and observation \citep{Beuther07}, and an initial temperature
$T_c=20~\textnormal{K}$.  The average density, initial radius and
free-fall time of the initial core is set by the initial volume
average core surface density
\begin{equation}
\Sigma = \frac{M}{\pi r_c^2}
\end{equation}
as
\begin{equation}
\rho_c = \sqrt{\frac{9 \pi \Sigma^3}{16 M}} \label{surfacescale}
\end{equation}
\begin{equation}
r_c = \sqrt{\frac{M}{\pi \Sigma}}
\end{equation}
and
\begin{equation}
t_\textsubscript{ff} = \left[\frac{\pi M}{64 G^2 \Sigma^3}\right]^{1/4}.\label{tff}
\end{equation}
The initial core is placed in the center of a cubical simulation
domain spanning $4$ times the core radius i.e., ($L_\textsubscript{domain}=4
r_c$) on each side, so that no part of the initial cloud except gas
that is entrained into protostellar winds ever approaches the
boundary.  The initial core is immersed into a uniform ambient
environment with density that is $0.01$ times that at the edge of the
initial core.  Pressure balance between the core and environment is
maintained by setting the temperature of the ambient gas to
$100$ times that at the edge of the initial core, $T_\textsubscript{
amb}=2000~\textnormal{K}$.  The Planck mean opacity of the ambient gas is
set to zero to ensure that it does not cool or radiatively heat the
core.  The cores are initialized with a turbulent velocity field
chosen to put them in approximate balance between gravity and
turbulent motions.  Three Gaussian random fields are generated with
power spectrum $P(k) \propto k^{-2}$ for the three velocity
components, each normalized to have an integrated norm of unity over
the full spectral range sampled. We set the initial velocity in every
cell equal to the components of the Gaussian random field times the
one dimensional velocity dispersion,
\begin{equation}
\sigma_v = \sqrt{\frac{GM}{2(k_\rho-1)r_c}} = \left[\frac{G^2 M \pi \Sigma}{4(k_\rho-1)^2}\right]^{1/4}\label{sigmav},
\end{equation}
corresponding to the velocity at the surface of a singular polytropic
sphere \citep{McKeeTan}.  The virial parameter $\alpha_\textsubscript{vir} = 5
\sigma_v^2 GM/r_c$ \citep{Bertoldi} is 5 for $k_\rho = 3/2$, somewhat
larger than the value of $15/4$ in hydrostatic equilibrium
\citep{McKeeTan}.  The kinetic energy is therefore initially larger
than the gravitational energy, but $\alpha_\textsubscript{vir}$ decreases with
time due to the decay of the turbulence.  The initial radiation energy
density is set to the value for a black-body radiation field with
radiation temperature $T_r=20~\textnormal{K}$ everywhere.

The earliest stages of high-mass star formation occur in infrared dark
clouds (IRDCs) \citep{rathborne07} which are detected in absorption
against the mid-infrared galactic background \citep{perault,egan}.
Observations of IRDCs indicate a range of surface density in star
forming regions from more tenuous sources with $\Sigma \sim 0.1
~\textnormal{g cm}^{-2}$, to more typical galactic star formation
conditions with $\Sigma \sim 1 ~\textnormal{g cm}^{-2}$, to $\Sigma \sim
10 ~\textnormal{g cm}^{-2}$ \citep{Beuther02b,rathborne,lopez} or more in
extragalactic super-clusters \citep{Turner,McCrady}.  Table
\ref{CommonParameters} summarizes the parameters for each of the four
computational models presented in this work.  The initial conditions
for these models have been chosen to study the collapse of galactic
IRDCs with high but not atypical mass and surface density.  Each of
the initial simulation core states are rescaled versions of one
another with identical density structure, virial ratio, velocity field
and comparable peak resolution in every run.  We have compared each of
the simulations at equivalent time in units of the free-fall time of
the initial cores.  The homology between the runs is broken only by
the presence or absence of outflows and by radiative effects.  This
choice of model parameters therefore probes the surface density
dependence of radiative feedback effects, and isolates the effects of
protostellar outflows by holding all other parameters constant as they
are turned on and off.

\begin{deluxetable}{l r r r r r}
  \tablewidth{0pt}
  \tablecaption{Simulation Parameters.}\label{CommonParameters}
    \startdata
    \tableline
    $\Sigma~(\textnormal{g cm}^{-2})$ & 1.0 & 2.0 & 2.0 & 10.0 \\
    wind & on & on & off & on \\
    $M~(M_\sol)$ & 300 & 300 & 300 & 300 \\
    $r_c~(\textnormal{pc})$ & 0.141 & 0.100 & 0.100 & 0.0447 \\
    $\bar{n}_H~(\textnormal{cm}^{-3})$ & $7.3\times10^5$ & $2.1\times10^6$ & $2.1\times10^6$ & $2.3\times10^7$ \\
    $\sigma_v/c_s$ & 8.80 & 10.5 & 10.5 & 15.6 \\
    $t_\textsubscript{ff}~(\textnormal{kyr})$ & 50.7 & 30.2 & 30.2 & 9.02 \\
    $L_\textsubscript{domain}~(\textnormal{pc})$ & 0.565 & 0.400 & 0.400 & 0.179 \\
    $N_0$ & 128 & 192 & 192 & 192 \\
    max level & 5 & 4 & 4 & 3 \\
    $\Delta x_L~(\textnormal{AU})$ & 28.4 & 26.8 & 26.8 & 24.0 \\
    \enddata
    \tablewidth{\textwidth}
    \tablecomments{
      Row 1: initial core surface density; 
      Row 2: initial core mass; 
      Row 3: initial core radius; 
      Row 4: initial core velocity dispersion; 
      Row 5: core free fall time; 
      Row 6: linear size of the computational domain,; 
      Row 7: number of cells per linear dimension on the coarsest level; 
      Row 8: maximum refinement level; 
      Row 9: computational resolution on the finest AMR level.
    }
\end{deluxetable}

We have largely followed the approach set forth by \cite{krumholz10}
in choosing the parameters for the numerical simulations considered
here.  However, the simulations in this work differ from the earlier
work in several ways.  First, these simulations use an initial core
mass of $300~\msun$ instead of $100~\msun$.  This choice was motivated
by our desire to study the evolution of high-mass star systems and our
expectation that protostellar outflows, which were not considered in
the earlier models, will eject a significant fraction of the initial
core.  Secondly, the lowest initial surface density is $\Sigma = 1.0~
\textnormal{g cm}^{-2}$ instead of $\Sigma = 0.1~ \textnormal{g cm}^{-2}$ as
used in the earlier work.  This choice is largely motivated by
computational constraints. The high flow speeds present in runs with
outflows necessitate smaller numerical time steps than for non-outflow
runs and thus increase the computational cost. A simulation with
$\Sigma = 0.1~\textnormal{g cm}^{-2}$ would be particularly expensive
due to its long free-fall time and the need to advance to these later
times with numerical time steps that are limited by the cell crossing
time of outflow-ejected gas.

\subsection{Refinement and Boundary Conditions}
The AMR capabilities of the code track the collapsing cores in three
dimensions to grid scales of $\Delta x_L \sim 25~\textnormal{AU}$ on the
finest AMR level.  This resolution is achieved by discretizing the
physical domain on the coarse onto a base grid of $N_0^3$ cells.  The
placement of finer level grids up to the finest level $L$ was
determined by the refinement criteria that any gas denser than one
half the density at the edge of the initial core be refined by at
least one level.  Further refinement is also triggered wherever the
local Jeans number, $J=\sqrt{G \rho \Delta x^2 / (\pi c_s^2)}$
\citep{Truelove}, exceeds $0.125$, where $\Delta x$ is the
computational cell width on the coarser level, or wherever the local
gradient of the radiation energy density $|\grad E_\textsubscript{rad}| \Delta x
/ E_\textsubscript{rad}$ exceeds $0.1$.

The simulations use a zero velocity gradient outflow boundary
condition for the hydrodynamics and Marshak boundary conditions for
the radiation energy density with radiation flux at the edge of the
computational domain for a $20~\textnormal{K}$ background radiation field.
The gravitational potential is specified at the edge of the domain as
the sum of the multipole moments of the mass distribution in the
computational volume as a function of time up to the quadrupole term.
We adopt an equation of state with $\gamma=5/3$, appropriate for gas
too cool for molecular hydrogen to be rotationally excited, but this
choice is essentially irrelevant because the gas temperature is set
almost purely by radiative effects.

\subsection{Optical Properties and Equation of State}
\begin{figure}[tpb]
\begin{center}
\includegraphics[clip=true,width=0.49\textwidth]{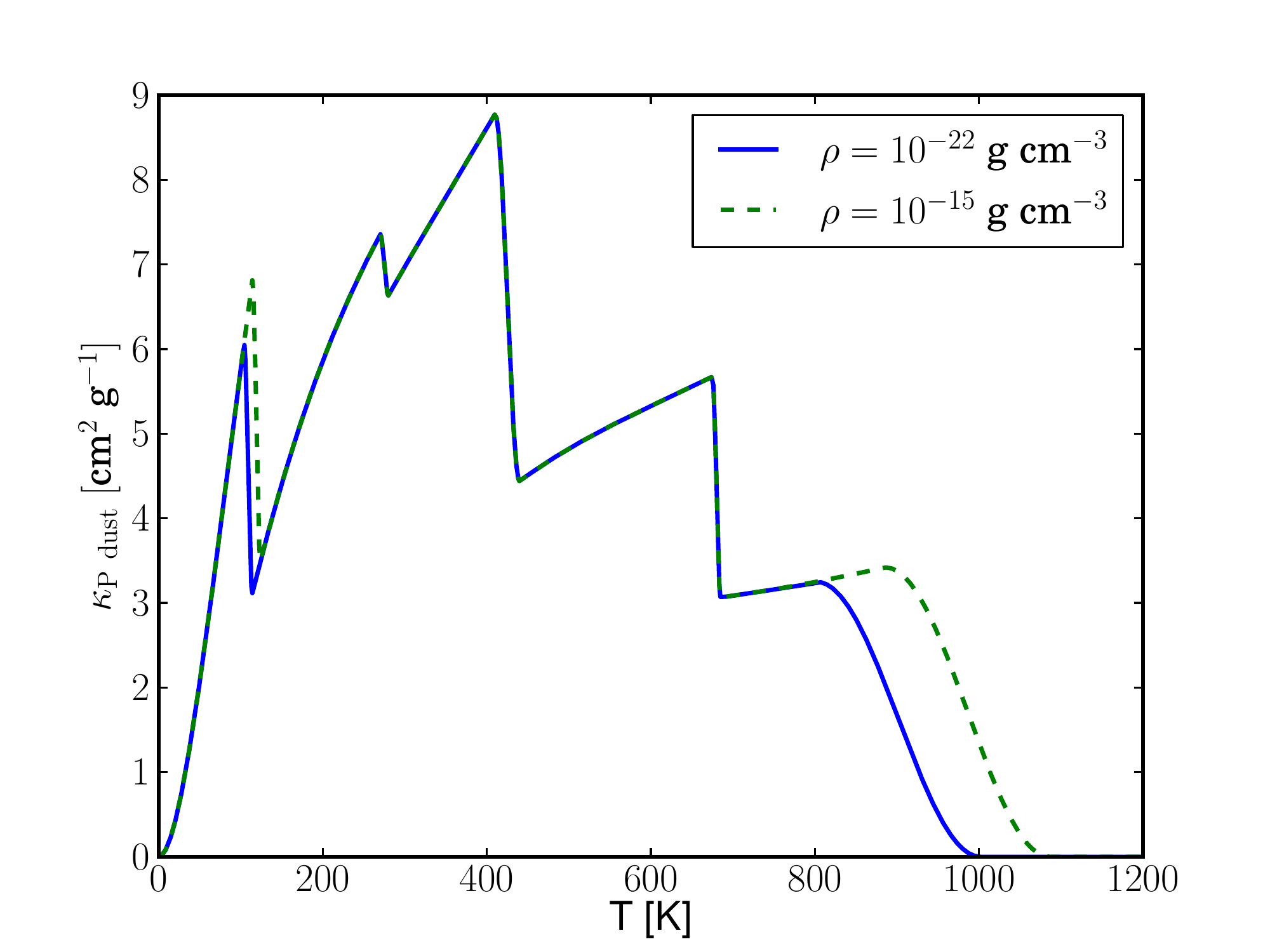}
\includegraphics[clip=true,width=0.49\textwidth]{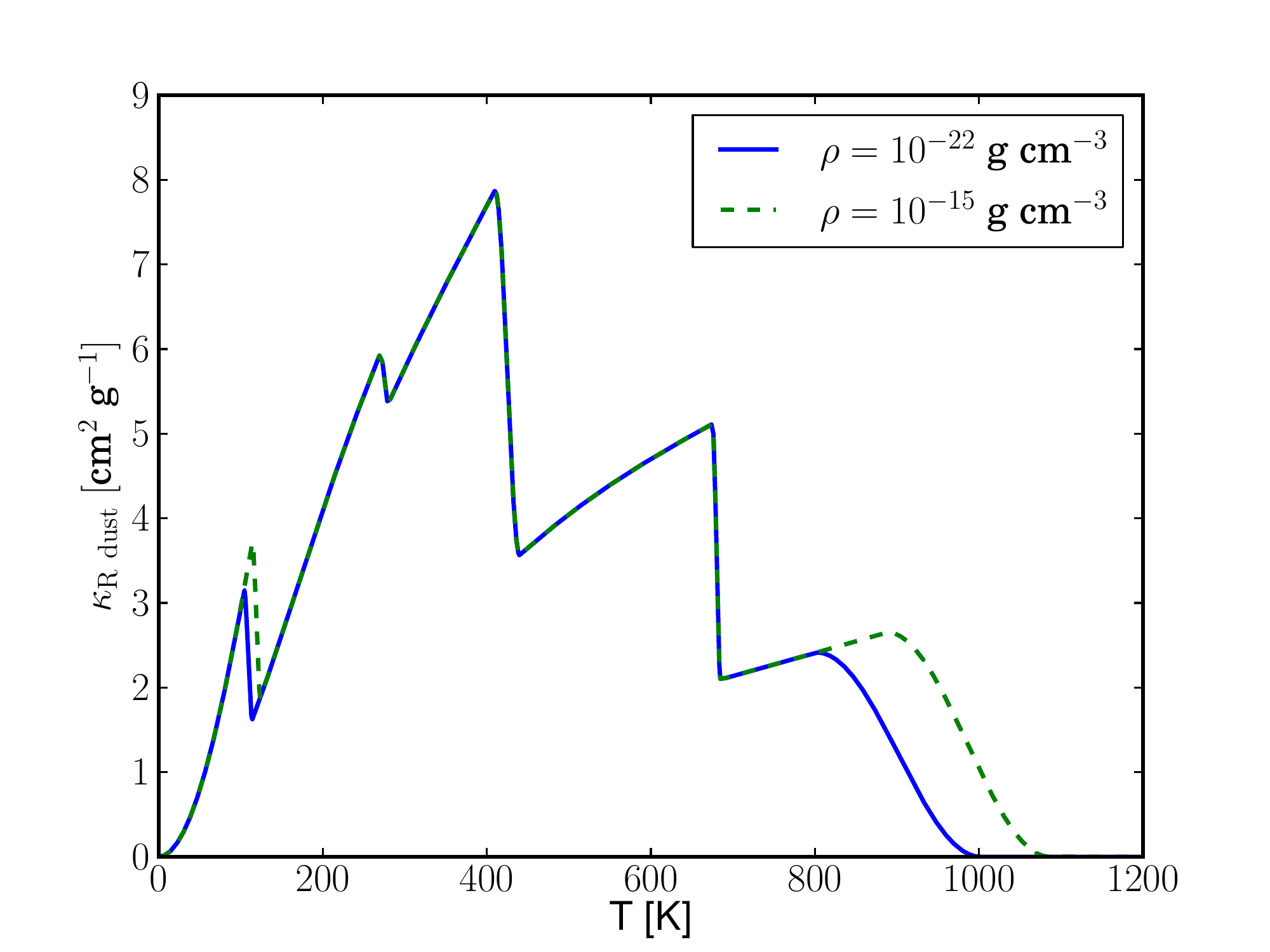}
\end{center}
\caption{Dust opacity model \citep{Semenov}. Left: Planck mean dust
  opacity as a function of gas temperature. Right: Rosseland mean dust
  opacity. \label{f1}}
\end{figure}

The radiation transport is handled by a frequency-integrated flux
limited diffusion approximation.  We use the Planck and Rosseland mean
dust opacities, $\kappa_\textsubscript{P}$ and
$\kappa_\textsubscript{R}$ respectively, of the \cite{Semenov} iron
normal, composite aggregates model plotted in figure \ref{f1}.  The
simulations with protostellar winds introduce strong heating behind
wind-driven shocks.  When the thermal energy of the gas exceeds that
of a gas with molecular weight $0.6 m_p$ and temperature
$10^4~\textnormal{K}$, we treat the gas as fully ionized with
Rosseland opacity $\kappa_\textsubscript{R} = 0.32$
$\textnormal{cm}^2~\textnormal{g}^{-1}$, the value for Thompson
scattering at solar metallicity.  Our gray flux limited diffusion
approximation cannot adequately represent the collisionally-excited
line cooling processes that dominates at temperatures above the dust
destruction temperature. However, it is critical to include them in
order to ensure that shocked gas is able to cool. We therefore leave
$\kappa_\textsubscript{P} = 10^{-2}$ cm$^2$ g$^{-1}$ for this gas,
ensuring that it does not interact strongly with the ambient radiation
field, but we also implement an approximate line cooling function
$\lambda(T)$ to remove energy from gas above the dust destruction
temperature and transfer it to the radiation field. We take
$\lambda(T)$ from the function shown in figure 1 of Cunningham et
al. (2006). In each time step, before we perform our ordinary flux
limited diffusion radiation solve, we update the gas and radiation
energy densities by implicitly solving the operator split-system
\begin{eqnarray}
\frac{d \rho e}{d t} &=& -(\rho/\mu)^2 \lambda(T) \\
\frac{d E}{d t} &=& (\rho/\mu)^2 \lambda(T)
\end{eqnarray}
using the LSODE \citep{LSODE} Gear-type solver.  In the above system,
$e$ is the specific thermal energy density of the gas, $E$ is the
radiation energy density, $\mu$ is the mean molecular weight and $T$
is the gas temperature appropriate for a solar metallicity ionized gas
mixture.

%
%

\subsection{Protostellar Wind Model} \label{wind}
The ORION code includes a ``star particle'' algorithm to handle the
formation of protostars \citep{KrumholzSinks,KrumholzStars}.  This
algorithm provides for the creation of sub-grid star particles in
those cells of the computation that become poised for gravitational
collapse to spatial scales smaller than those that can be captured on
the computational grid without spurious fragmentation
\citep{Truelove97}.  The luminosity, radius and burning state of the
star particle is advanced with the simulation according to the
protostellar evolution model of \cite{McKeeTan} as updated by
\cite{Offner}.  The protostellar evolution model takes as input the
mass and accretion history of the star as determined by the simulation, and
as output predicts a protostar's radius and luminosity at any given time.
The protostellar luminosity prescribed by this model enters the simulation
as a source term in the radiation energy density equation, and the
protostellar radius is used to compute the Keplerian velocity at the stellar
surface, which affects outflows as described below.  The protostellar
luminosity prescribed by this model enters the simulation as a source
term in the radiation energy density equation.

The ORION star particle algorithm has been enhanced for this work to
include the driving of bipolar outflows.  Our outflow model is
specified by the dimensionless parameters $f_w$ and $f_v$, which set a
wind launch speed as a fraction $f_v$ of the Kepler speed at the
stellar surface and a mass flux that is a fraction $f_w$ of the rate
of accretion onto the star or, equivalently, a fraction $f_w/(1+f_w)$
of the total mass that is either accreted onto the star or ejected in
the wind.  Since we are interested in the large scale impact of the
protostellar winds, we assume that the wind is injected over a range
of radii determined by a function $\chi_w(|\vec{r}|)$ with an angular
dependence given by $\bar \xi(\theta_i)$; explicit forms for these
functions are given below.  The wind driving is imposed by operator
split source terms in the gas density, momentum density and energy
density equations with
\begin{eqnarray}
\left. \frac{d \rho}{d t} \right|_s         &=& -\dot{M}_{w,i}~\chi_w(|\vec{r}_i|)~\bar{\xi}(\theta_i), \\
\left. \frac{d \rho \vec{v}}{d t} \right|_s &=& -f_v v_{k,i}~\dot{M}_{w,i}~\chi_w(|\vec{r}_i|)~\bar{\xi}(\theta_i)~\cdot \vec{\hat{r}}_i, \label{msx} \\ 
\left. \frac{d \rho e}{d t} \right|_s       &=& -\dot{M}_{w,i}~\chi_w(|\vec{r}_i|)~\bar{\xi}(\theta_i)~\frac{k_B T_w}{\mu (\gamma-1)},
\end{eqnarray}
where 
\begin{equation}
v_{k,i}=\sqrt{\frac{GM_i}{r_{*,i}}} \label{kepler}
\end{equation}
is the Keplerian speed at the surface of the star and $r_{*,i}$ is the
protostellar radius; as remarked above, we use the value of
$r_{*,i}$ given by the model of \cite{McKeeTan} as updated by
\cite{Offner}.  For the simulations presented here we have set the
wind-launched gas temperature as $T_w = 10^4~\textrm{K}$, appropriate
for an ionized wind.  The corresponding rate of particle mass growth,
particle wind mass ejection rate, acceleration, radial distance and
spatial inclination of the $i^{th}$ star are
\begin{eqnarray}
\dot{M}_i          &=&  \frac{1}{1+f_w} \dot{M}_{KKM04}, \label{sb} \\
\dot{M}_{w,i}       &=&  f_w \dot{M}_i = \frac{f_w}{1+f_w} \dot{M}_{KKM04}, \\
\vec{\dot{v}}_{i}  &=&  \vec{\dot{v}}_{KKM04}, \\
\vec{r}_i          &=& \vec{x}-\vec{x}_i, \\
\theta_i           &=& \textnormal{acos}( \hat{\vec{r}}_i \cdot \hat{\vec{j}}_i )
\end{eqnarray}
where $\vec{j}_i$ and $\vec{x}_i$ are the velocity and position of
the $i^{th}$ particle, $\vec{j}_i$, $\dot{M}_{KKM04}$ and $\vec{\dot{v}}_{KKM04}$
are the sink particle angular momentum, accretion rate and acceleration for
the case in which winds are absent as given by the algorithm of
\cite{KrumholzSinks}.  

Values of the parameters $f_w$ and $f_v$ can be both estimated from
theory and constrained by observations. Theoretically, the X-wind
\citep{Shu88} and disk wind \citep{Pelletier92} models predict $f_w
\sim 0.3$, $f_v \sim 1$ and $f_w \sim 0.1$, $f_v\sim 3$,
respectively. Both therefore suggest $f_w f_v \sim 0.3$.  The total
momentum $P_w$ carried by an observed outflow from a star of mass
$M_*$ is related to $f_w$ and $f_v$ by
\begin{equation}
f_w f_v = \frac{P_w}{v_k M_*}, \label{fwfv}
\end{equation}
where $v_k$ is the Keplerian velocity at the stellar surface.  The
peak of the stellar initial mass function is at $M_* \approx 0.2
\msun$ \cite{Chabrier}, and such stars typically have radii $\sim 3
\rsun$ during their main accretion phases (e.g.\ model mC5H of
\cite{Hosokawa}), so typical values of $v_k$ are $\sim 100~\textrm{km
s}^{-1}$. This value of $v_k$ together with $f_w f_v \sim 0.3$ implies
a net wind momentum flux of $\sim 30~\textrm{km s}^{-1}$ per $\msun$
of stars formed.

A number of surveys have used measurements of $P_w$ and estimates of
$M_*$ and $v_s$, together with the relation given in equation
(\ref{fwfv}), to constrain $f_w f_v$. Surveying the literature
available as of 2000, \cite{Richer} estimate $f_w f_v \sim 0.3$.  More
recent observational surveys of several nearby low-mass star forming
regions indicate typical outflow momenta of $\sim 0.2$ to $\sim
3.0~\msun \textrm{ km s}^{-1}$ \citep{Maury,Arce,Curtis,Ginsburg}.
The physical properties of the driving sources of most of the surveyed
outflows are not very well constrained.  However, if we assume that
the typical source has accreted half of its final mass $M_* \sim 0.1
\msun$ and has radius $r_* \sim 3 \rsun$ then equations (\ref{kepler})
\& (\ref{fwfv}) can be use to extract a range of outflow momentum
parameters based on the results of these surveys of $0.025 \lesssim
f_w f_v \lesssim 0.38$.

Observationally, $f_w$ and $f_v$ can be better constrained from
sources where observational measurements exist for both net outflow
momentum and the net mass accreted onto the protostar $M_*$.
\cite{Curtis} have surveyed the outflow momentum in the young cluster
NGC1333 and \cite{SandellKnee} have estimated the mass of warm dusty
gas in the collapsing envelope around the deeply embedded protostars
that drive several of these outflows.  In table \ref{toutflows} we
list the intersection of those sources that have both dust mass
measurements from \cite{SandellKnee} and outflow momentum measurements
by \cite{Curtis}, excluding a few sources that \cite{SandellKnee}
indicated as near the edge of their field of view with unreliable flux
densities, and excluding one tight binary source that could not be
separately resolved in that work (IRAS 4A).  We expect that the
envelope masses of early class 0 sources should be somewhat greater
than the mass of the embedded protostar and we expect that the mass of
later class I sources should exceed that of their envelope.  In the
absence of better mass constraints for the collection of class I and 0
protostars in the present sample we adopt the assumption $M_* \sim
M_{Dust}$ as a rough approximation.  We assume fiducial stellar radius
of $r_* = 3~\rsun$ following model mC5H of \cite{Hosokawa}.  From
these assumptions, we can constrain $f_w f_v$ from equations
(\ref{fwfv}) \& (\ref{kepler}). The datum indicate a range of outflow
launch parameters of $0.01 \lesssim f_w f_v \lesssim 0.15$, with no
strong statistical correlation of $f_w f_v$ with the source spectral
class.

\begin{deluxetable}{l r r r r r}
  \tablewidth{0pt}
  \tablecaption{NGC1333 protostellar outflow data.}\label{toutflows}
  \startdata
  \tableline
  Source & Class & $M_{Dust}$ & $P_w$ & $f_w f_v$ \\
  \tableline
  HRF42 & 0 & 0.49 & 0.058 & 6.7$\times10^{-4}$ \\
  HRF43 & I & 0.36 & 3.08 & 0.057 \\
  HRF44 & 0 & 0.35 & 3.17 & 0.061 \\
  HRF45 & I & 0.31 & 0.28 & 6.4$\times10^{-3}$ \\
  HRF46 & 0 & 0.1  & 0.44 & 0.055 \\
  HRF47 & 0 & 0.24 & 0.23 & 7.8$\times10^{-3}$ \\
  HRF54 & I & 0.3  & 0.10  & 2.4$\times10^{-3}$ \\
  HRF56 & I & 0.04 & 0.11 & 0.052 \\
  HRF62 & 0 & 0.32 & 0.23 & 0.0051 \\
  HRF63 & I & 0.08 & 0.07 & 0.011 \\
  HRF65 & 0 & 0.07 & 0.77 & 0.17 \\
  \tableline
    min   &   &      &      & 6.7$\times10^{-4}$ \\
    mean  &   &      &      & 0.039 \\
    max   &   &      &      & 0.17 \\
  \enddata
  \tablewidth{\textwidth}
  \tablecomments{
    Column 1: \cite{Hatchell} source number; 
    Column 2: Spectral Class \citep{Hatchell}; 
    Column 3: Progenitor mass \citep{SandellKnee}; 
    Column 4: Progenitor mass from the table B2 in the online supplementary data to \citep{Curtis}; 
    Column 5: Implied outflow launch parameter assuming $v_w = 100~\textrm{km s}^{-1}$.
  }
\end{deluxetable}

Wind launch speeds represent the Courant time-step constraint in a
typical calculation, so large values of $f_v$ impose a particularly
onerous computational burden.  We therefore choose a wind mass to
stellar mass fraction on the high end of the theoretical guidance,
$f_w = 27\%$ and the wind velocity parameter of, $f_v = 1/3$.  This
yields a momentum flux injected by our wind model characterized by
$f_w f_v = 9\%$, which is toward the higher end of the observed range
of rates of momentum injection by outflows from the low mass sources
tabulated in table \ref{toutflows}.

The function
\begin{equation}
\chi_w(r) = \frac{1}{C_1}\left\{ \begin{array}{ll}
r^{-2} & \textnormal{if~} 4 \Delta x < r \leq 8 \Delta x \\ 
0 & \textnormal{otherwise} 
\end{array} \right. \label{kernel}
\end{equation}
is a normalized weighting kernel that determines the spatial scale
of the wind injection where $C_1$ is a normalization constant to the
weighting kernel.  The $C_1$ is computed numerically in the ORION code so
that numerical aliasing effects of the spherical wind injection region
into the Cartesian grid are accounted for exactly.

The remaining function $\bar{\xi}$ describes the angular distribution of the
wind mass and momentum flux at the point where it is injected into the
computational grid. We take this function from \cite{MatznerOutflows},
who find that the momentum distribution of prestellar outflow
injection asymptotically far away from the protostellar surface as a
function of the polar angle $\theta$ from the direction of the
protostar's rotation is given by
\begin{equation}
\xi(\theta,\theta_0) = \left[\ln \left(\frac{2}{\theta_0} \right) (\sin^2 \theta +\theta_0^2)\right]^{-1} \label{matz}
\end{equation}
where $\theta_0$ is the so called ``flattening parameter'' that sets
the opening angle of the wind.  In the case of low-mass stars,
\cite{MatznerOutflows} suggest a fiducial value of $\theta_0=0.01$ and
we adopt the same value here.  While stars of type B or later direct
momentum in a very well-collimated beam
\citep{Beuther02a,Beuther03,Beuther04}, O star winds are collimated
somewhat more weakly, possibly due to the effect of ionization
\citep{Beuther05}.  Since we do not include ionizing radiation in
these simulations, we do not attempt to model to model O star wind
broadening.

The large value of $\xi$ near $\theta=0$ in equation (\ref{matz}) requires
particular care in implementing this model in a numerical code.  We
implement the driving function by averaging $\xi$ over the polar angle
subtended by a grid cell $\Delta \theta = \atan(1/8)$ at the outer
radius of the weighting kernel (equation (\ref{kernel})) as
\begin{equation}
\bar{\xi}(\theta,\theta_0) = \frac{1}{C_2} \left\{ \begin{array}{ll}
\frac{1}{\Delta \theta}\int_{\theta-\Delta \theta /2}^{\theta+\Delta \theta /2} \frac{d\theta}{\sin^2 \theta +\theta_0^2} & \textnormal{if~}  |\sin(\frac{\pi}{2}-\theta)| \geq  \frac{\Delta x}{r}\\ 
0 & \textnormal{otherwise}. 
\end{array} \right. \label{avgkernel}
\end{equation}
Our choice to set $\bar{\xi}$ to zero for angles close to $\pi/2$ is
driven by numerical considerations. If we allow the outflow to be
injected into $4\pi$ steradians around the star, its mass and momentum
are sufficient to disrupt the early development of an equatorial
disk. This behavior is an artifact of the necessarily poor resolution
inside the wind launching region.  We do not resolve the disk scale
height, and this artificially puffs up the disk and reduces its mass
and momentum density, rendering it far easier for the outflow to
disrupt than it would be if its true scale height were resolved.  We
avoid this problem by reducing the outflow mass and momentum flux to
zero in an equatorial belt that is at least one cell thick, ensuring
that accreting material always has an uninterrupted path to the star.
We note that \cite{Schonke} have considered the effect of
  radiative and protostellar outflow feedback on the dynamics of the
  accretion disk in two dimensions at much higher resolution than the
  simulations considered here.  Their simulations indicate that
  feedback effects can alter the accretion rate onto the star on
  shorter timescales and smaller length scales than have been resolved
  in this study.  However, the emphasis of the present study is on the
  large scale radiative and feedback effects on the ambient core and
  we do not attempt to model this small-scale behavior.

The definite integral in equation (\ref{avgkernel}) evaluates to
%
\begin{eqnarray}
&&\int_{\theta-\Delta \theta /2}^{\theta+\Delta \theta /2} \frac{d\theta}{\sin^2 \theta +\theta_0^2} = \nonumber \\
&&\frac{1}{\theta_0 \sqrt{\theta_0^2+1}} 
\left[
      \atan \left(\frac{\sqrt{\theta_0^2+1} \tan \left(\theta+\frac{\Delta \theta}{2}\right)}{\theta_0}\right)-
      \atan \left(\frac{\sqrt{\theta_0^2+1} \tan \left(\theta-\frac{\Delta \theta}{2}\right)}{\theta_0}\right)
\right]
\end{eqnarray}
The normalization constant $C_2(\theta_0) = \int \xi(\theta,\theta_0)
\chi_w(r) d^3\vec{x}$ is also computed numerically to exactly account
for grid aliasing effects.  Neglecting the grid aliasing effect, we
find $C_2 = 8.165$ by numerical integration.  A second subtlety that
arises in the numerical implementation of the wind driving is that
care must be taken so that the momentum source terms impart exactly
zero net momentum onto the star particles and the gas in the
computational domain.  If the position of the particle is allowed to
vary continuously within its host cell the $\frac{1}{r^2}$ term in
equation (\ref{kernel}) may lead to an asymmetric driving.  To
overcome this problem, we bring the wind driving into symmetry with
the numerical grid by rounding the particle position to the nearest
half-integer multiple of the grid width
\begin{equation}
\vec{x_i} \leftarrow 2 \Delta \vec{x}~\textnormal{nint}\left[ \frac{\vec{x_i}}{2 \delta \vec{x}} \right]
\end{equation}
before computing the source terms (equations (\ref{sb}-\ref{msx})) where
$\textnormal{nint}$ is the ``nearest integer'' function.

The wind momentum injection $\bar{\xi}$ is significantly broadened at
polar latitudes relative to the analytic prescription for
$\xi$. Consequently less momentum is injected near the equator to
satisfy the normalization constraint.  To facilitate comparison of our
numerical models with the analytic predictions of \cite{MatznerCores}
in \S\ref{sfe} it is convenient to define the effective numerical
flattening parameter $\theta_{0,{\rm eff}}$ over a range of angles
that separate the outflowing gas from ambient gas:
\begin{equation}
  \xi(\theta,\theta_0=\theta_{0,{\rm eff}})= \bar{\xi}(\theta,\theta_0=0.01).
\end{equation}
We find that the above expression holds to with $10\%$ for
$\theta>10^\circ$ with $\theta_{0,{\rm eff}}=5.75 \times 10^{-4}$.

\section{Results} \label{results}
\subsection{Large-Scale \& Outflow Morphology} \label{largescale}
The left column of figures \ref{f2} through \ref{f5} show the large
scale evolution of each simulation from $t=0.2 t_\textsubscript{ff}$
to $t=0.8 t_\textsubscript{ff}$ in increments of $0.2
t_\textsubscript{ff}$ .  These plots show slices of density with the
color-mapping scaled by $\Sigma^{3/2}$ following the scaling given in
equation (\ref{surfacescale}).  The spatial scale shown in each plot
scaled by the initial core radius with each showing an area $(2.5
r_c)^2$.  The slices are centered on the position of the primary
protostar and oriented so that the angular momentum vector of the
protostar is upwardly oriented on the page.  As expected, once the
scaling of cloud radius and surface density is accounted for, the
regions that have not been penetrated by the outflow bow shocks
collapse in a homologous manner with little dependence on the initial
surface density.  The protostellar outflows evacuate a shock-bounded
cavity through the initial cores and the outflow cavities are the
prominent features in the density slices in the left column by $t \sim
0.3 t_\textsubscript{ff}$ in the lower surface density cases and $t
\sim 0.4 t_\textsubscript{ff}$ in the high surface density case.  The
propagation speed of the outflow bow shocks through the core and the
width of the outflow cavities show a strong dependence on initial
surface density.  The lower surface density cores show significantly
greater disruption due to the protostellar outflow feedback.  This
effect is due to 1) greater mechanical luminosity of the protostellar
winds at lower surface density (an effect that we discuss in detail in
\S\ref{properties} and 2) lower turbulent $(\sigma_v^2 \propto
\Sigma^{1/2})$ and thermal pressures $(P_c \propto \rho_c T_c \sim
\Sigma^{3/2})$ in the ambient cores which act to confine the
propagation of the outflows.  We note that the outflow-evacuated
cavities in the cases with outflows emerge from the initial core
toward the left side of the density slices in figure \ref{f5}.  The
reason for this is that the primary star particle retains the momentum
of the material that it accretes and therefore the star drifts away
from the center of mass of the initial core with time (see
\S\ref{properties}).  The outflowing material therefore emerges first
from the thinnest side of the initial cloud, relative to the position
of the primary star.

\subsection{Fragmentation \& Star Formation} \label{smallscale}
\begin{figure}[tpb]
\begin{center}
\includegraphics[clip=true,width=0.9\textwidth]{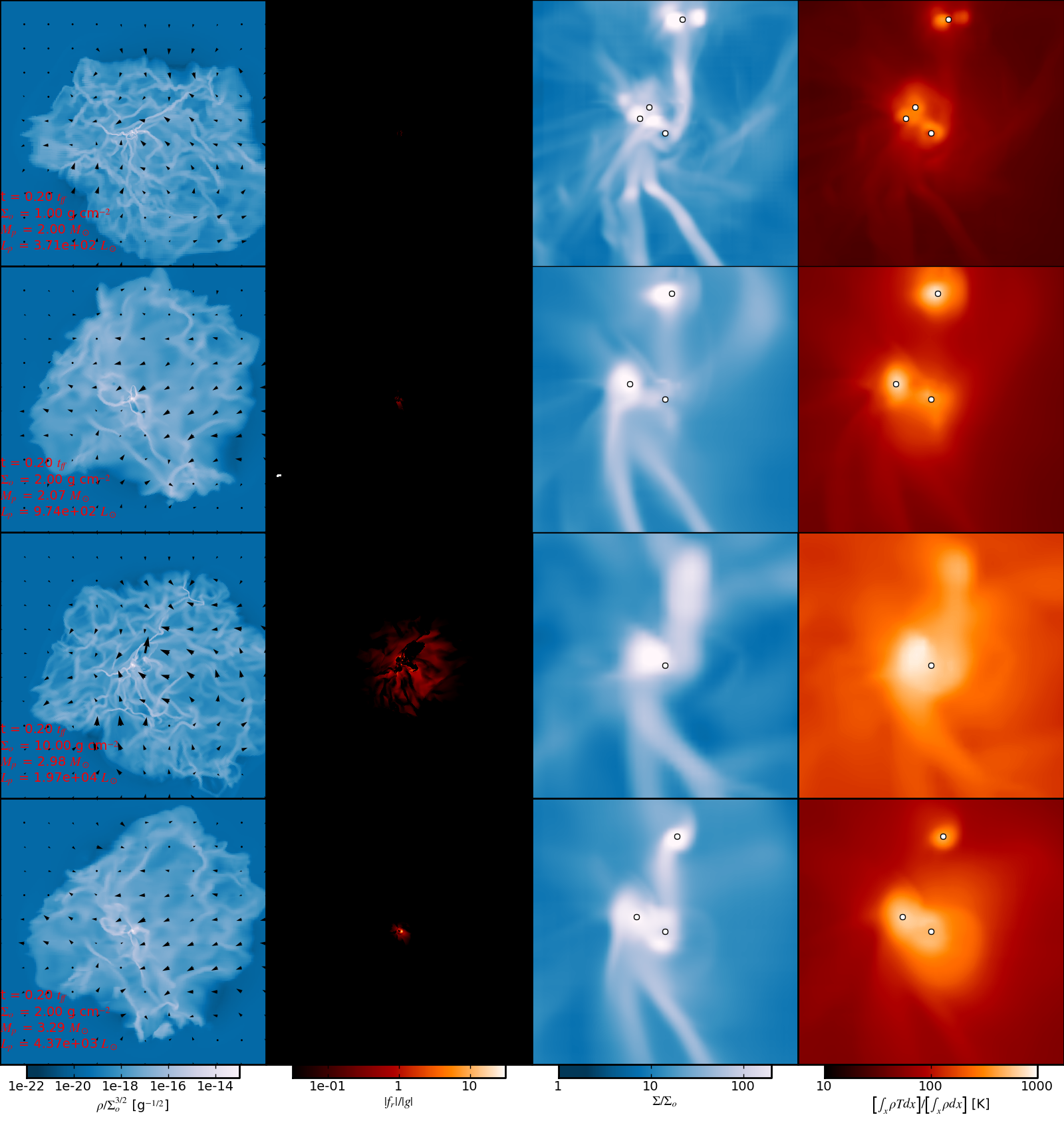}
\end{center}
\caption{Simulation graphics at $t=0.2 t_\textsubscript{ff}$ for the
  parameters $\Sigma=1.0\textnormal{g cm}^{-2}$,
  $\Sigma=2.0\textnormal{g cm}^{-2}$, $\Sigma=10.0\textnormal{g
  cm}^{-2}$, and $\Sigma=2.0\textnormal{g cm}^{-2}$ (without winds)
  from top to bottom.  First column: $\rho / \Sigma^{3/2}$ on a $(2.5
  r_c)^2$ plane oriented so that the outflow launch direction lies in
  the plane of the image, pointing toward the top of the page. 
  Black arrows indicate the velocity field.  An arrow with length
  equal to $1/8$ of the plot width indicates a flow speed of
  $100~\textrm{km s}^{-1}$, and arrow lengths scale as
  $\sqrt{|\vec{v}|}$. Second Column: Ratio of the radiation force
  magnitude to gravitational force magnitude. Third Column: Column
  density on a $(0.1 r_c)^2$ plane aligned with the cardinal axes of
  the simulation, oriented so that the primary protostellar outflow
  direction is as close as possible to pointing vertically out of the
  page.  Fourth Column: Mass-weighted radiation temperature projected
  in the same manner as the surface density in the third column.  All
  plots are centered on the projected position of the primary star.
  White markers indicate star particles. \label{f2}}
\end{figure}

\begin{figure}[tpb]
\begin{center}
\includegraphics[clip=true,width=0.9\textwidth]{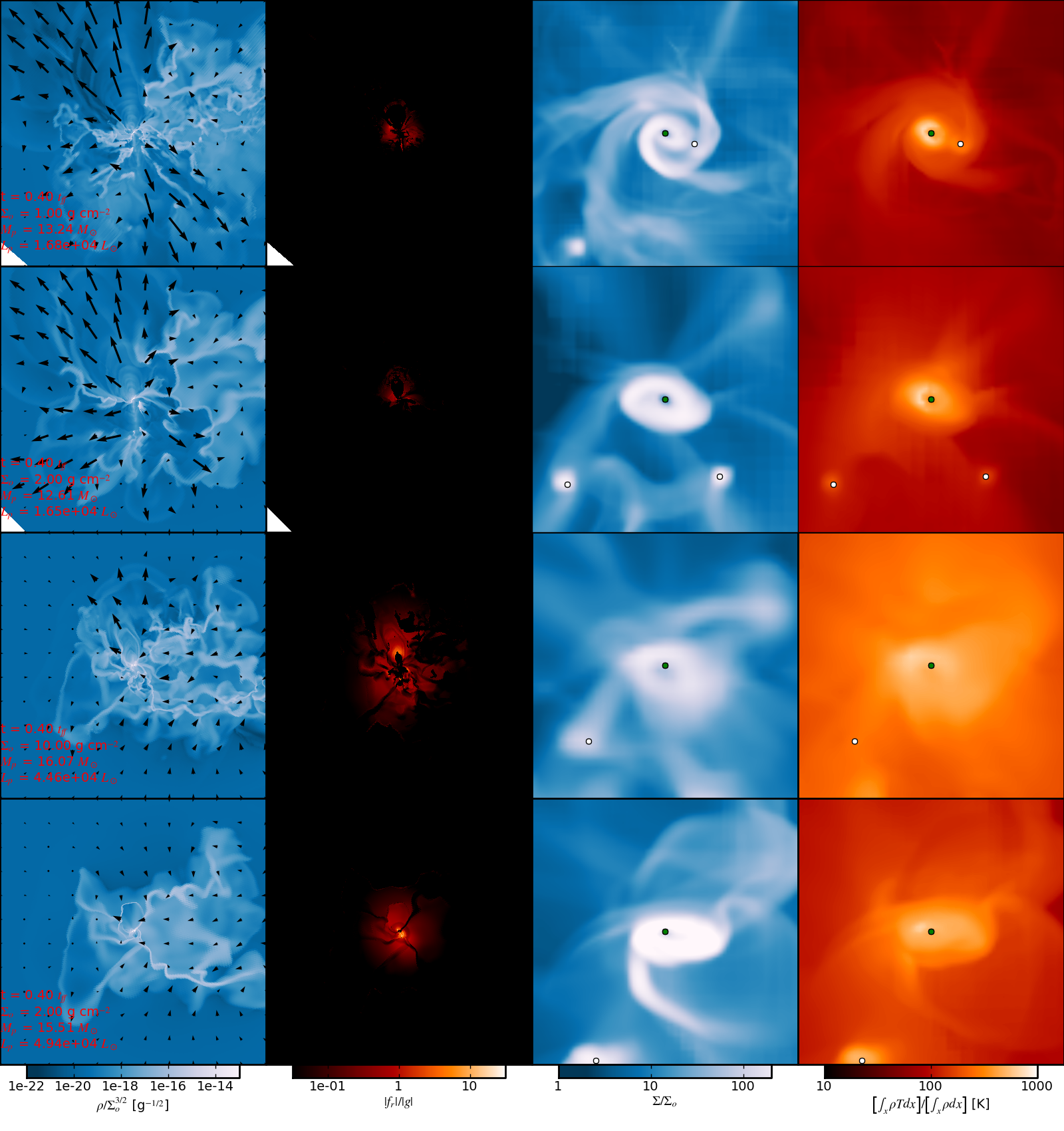}
\end{center}
\caption{Same as figure \ref{f2} but at $t=0.4 t_\textsubscript{ff}$. \label{f3}}
\end{figure}

\begin{figure}[tpb]
\begin{center}
\includegraphics[clip=true,width=0.9\textwidth]{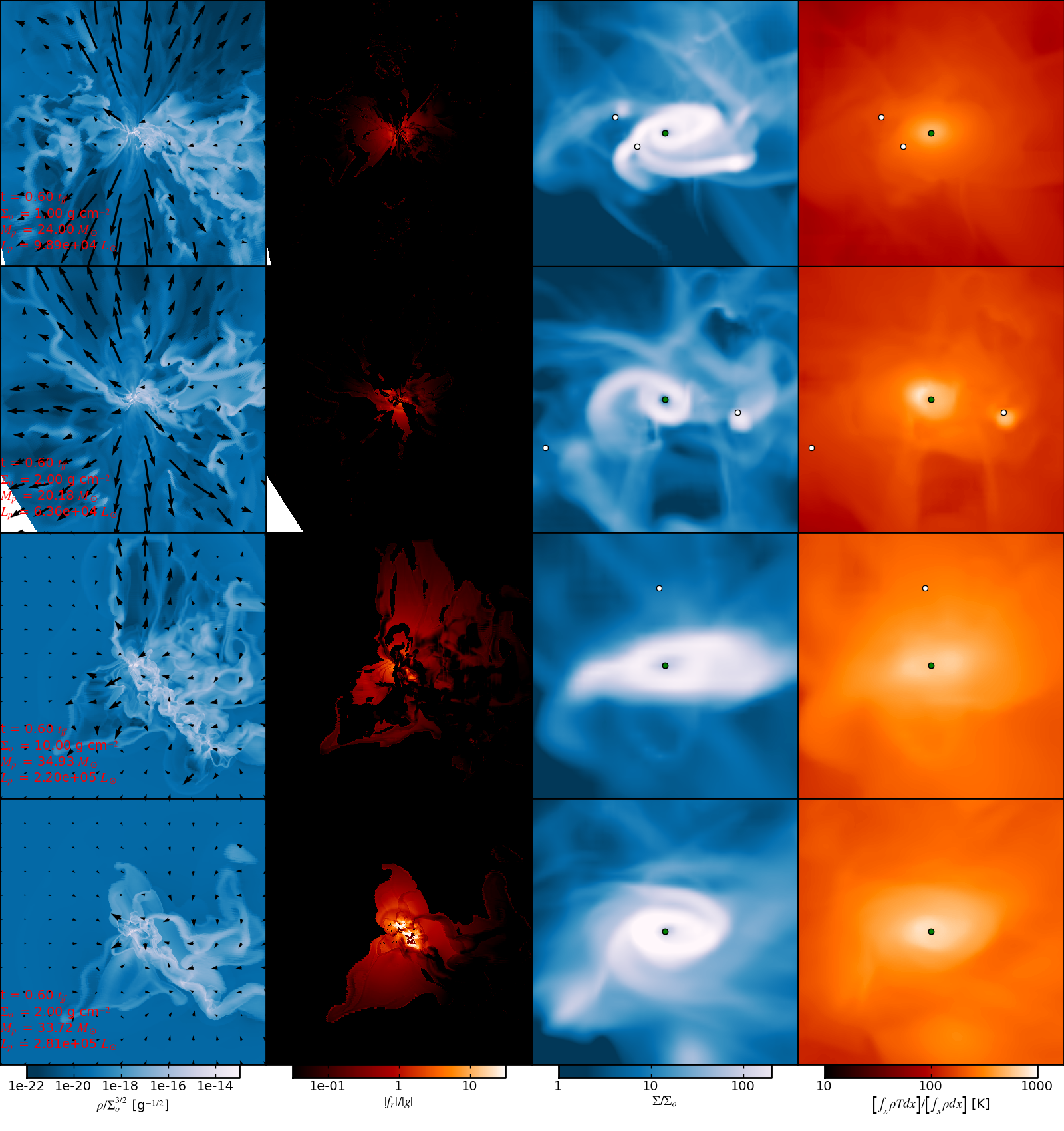}
\end{center}
\caption{Same as figure \ref{f2} but at $t=0.6 t_\textsubscript{ff}$. \label{f4}}
\end{figure}

\begin{figure}[tpb]
\begin{center}
\includegraphics[clip=true,width=0.9\textwidth]{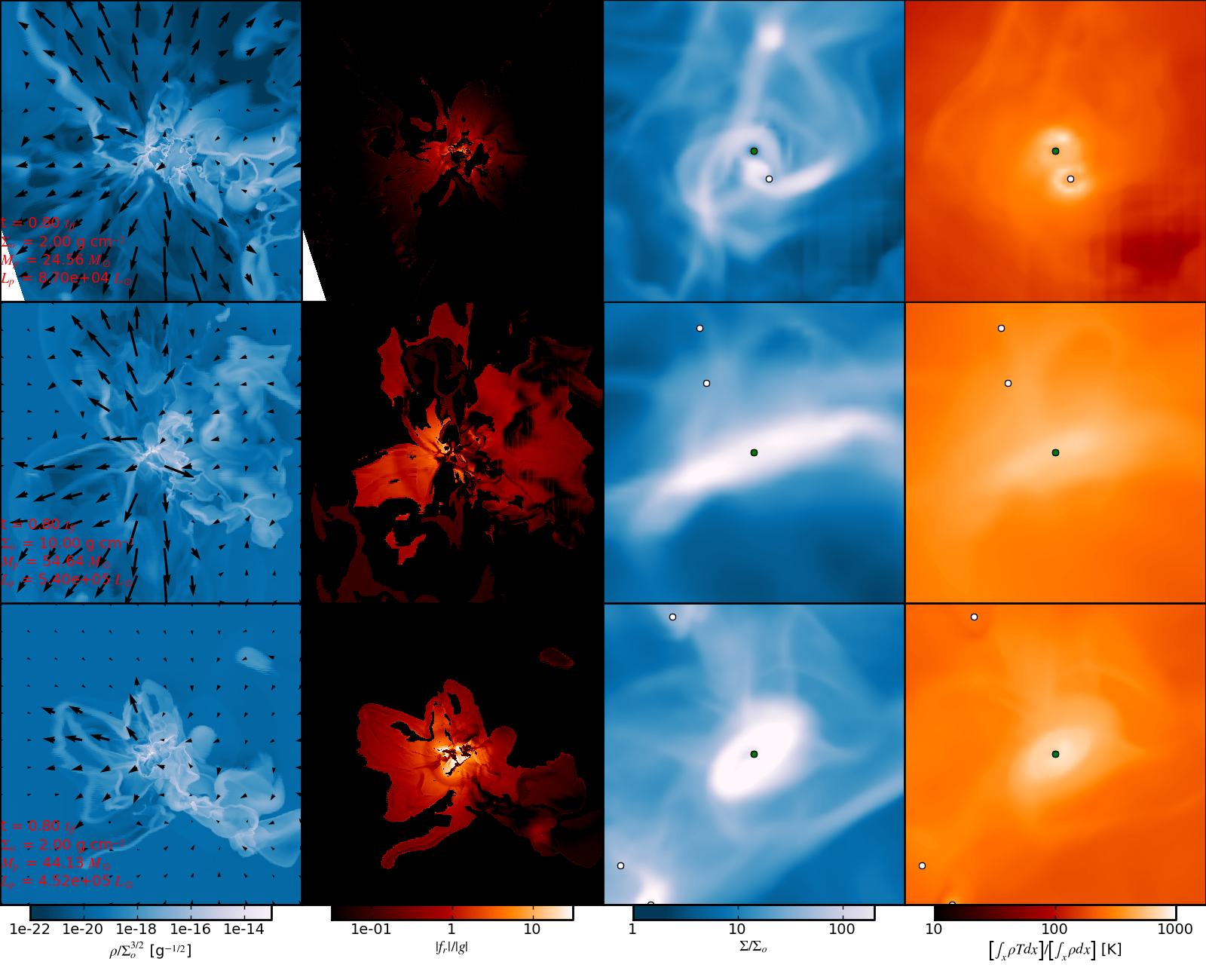}
\end{center}
\caption{Same as figure \ref{f2} but at $t=0.8 t_\textsubscript{ff}$. Only the
  surface density parameter cases of $\Sigma=2.0\textnormal{g cm}^{-2}$,
  $\Sigma=10.0\textnormal{g cm}^{-2}$, and $\Sigma=2.0\textnormal{g cm}^{-2}$
  (without winds) were run to this time.  \label{f5}}
\end{figure}

The third column of figures \ref{f2} through \ref{f5} show the small
scale evolution of the surface density.  Each of the plots are scaled
by the initial surface density $\Sigma$ and the initial core radius,
with each showing an area $(0.1 r_c)^2$ centered on the position of
the primary star.  Deviations from the the homologous scaling with
surface density exhibited on larger scales appear by $t=0.2
t_\textsubscript{ ff}$.  By $t=0.4 t_\textsubscript{ff}$ in figure
\ref{f3} and $t=0.6 t_\textsubscript{ff}$ in figure \ref{f4}, notable
differences in the disk around the primary protostar emerge.
Progressing from the third row, showing the case with highest surface
density, to the top row, showing the case of lowest surface density we
note an increasing tendency of the disk around the primary star to
fragment and generate spiral arms characteristic of a Toomre-unstable
disk \citep{Toomre}.  In the simulations with lower surface densities
in the top two rows of figure \ref{f4}) we note a prominent increase
in disk fragmentation in terms of the multiplicity of star particles
and the presence of distinctly separate accretion disks around the
primary and secondary protostars in comparison to the highest surface
density case in the third row.

The trend toward reduced disk stability and enhanced fragmentation
with lower initial surface density is consistent with earlier analytic
\citep{Krumholz08} and numerical \citep{krumholz10} works that predict
a threshold core surface density for sufficient radiative heating to
inhibit disk fragmentation of $\Sigma \sim 1.0~\textnormal{g
  cm}^{-2}$, as well as with observational data from infrared
star-forming clouds that are consistent with this prediction
\citep{lopez}.  We will show in \S\ref{focusing} that protostellar
outflows provide a mechanism for focusing radiative feedback in the
poleward directions, away from the the infalling disk in the midplane,
as predicted by \cite{KrumholzOutflows}.  Protostellar outflows should
therefore raise the surface density threshold for high-mass star
formation.  The simulations presented here do not survey sufficiently
low surface densities to quantify this effect, due to the
computational cost of simulating protostellar winds inside low surface
density cores with commensurately longer free fall times.
Furthermore, we expect that the relative importance of this effect may
depend on magnetic fields, which our simulations neglect, and their
role in confining the outflow cavity \citep{Hennebelle}.  We therefore
defer more precise quantification of the effect of outflows on the
minimum surface density threshold for high-mass star formation to
future work that will include magnetic fields and survey lower surface
density cores.

We can isolate the effect of protostellar outflows on the small-scale
evolution by comparing the second and fourth rows of figures \ref{f2}
through \ref{f5} which show the results of our numerical experiments
with and without protostellar outflow ejection, both with the same
initial surface density of $\Sigma = 2.0~\textnormal{g cm}^{-2}$.  By
$t=0.4 t_\textsubscript{ff}$ enhanced radiation trapping in the case with
outflows has lead to substantially warmer circumstellar gas in
comparison to the case without outflows.  The temperature structure in
the $\Sigma = 2.0~\textnormal{g cm}^{-2}$ case {\it without} protostellar
outflows in the fourth row is more similar to the $\Sigma =
10.0~\textnormal{g cm}^{-2}$ case {\it with} outflows in the third row
than to the case at the same surface density with winds in the second
row.  As we will show in \S\ref{focusing}, protostellar outflow
cavities carve a path of reduced optical depth through the initial
core that channel radiative flux away from the center of the core.
This escape of radiative energy reduces the efficacy of radiative
heating in the central regions of the collapsing core.

Enhanced disk fragmentation associated with outflows and decreasing
surface density is due to the reduced effectiveness of radiative
heating.  The right columns of figures \ref{f2} through \ref{f5} show
the mass-weighted, column-averaged radiation temperature $T_r$ over the
same regions as the column density projections in the center column.
We note that the dense infalling gas is strongly coupled to the
radiation field, so that the gas temperature $T \approx T_r$.  Only
the tenuous outflow-evacuated regions achieve sufficient post-shock
temperatures to break the radiation-gas coupling by increasing the gas
temperature beyond the dust destruction temperature.  The
column-averaged radiation temperature plots therefore allow us to
probe the temperature structure of the dense infalling gas without
confusion from the projection of shock heated layers along the
surfaces of the outflow cavity.  The temperature structure of the gas
on small scales shows even stronger dependence on the initial surface
density than the column density.  We note enhanced temperature with
increasing surface density as early as $t=0.2 t_\textsubscript{ff}$ when no
stars are producing significant power via nuclear burning.  The
dependence of temperature on surface density at these times is solely
due to the factors pointed out by \cite{Krumholz08}; (1) Higher
surface density cores have higher accretion rates, thereby generating
higher accretion luminosity and (2) higher surface density cores have
higher optical depth to more effectively trap radiation.  The trend
toward increasing gas temperature with increasing surface density is
also present at all later times, due to additional
radiative output from nuclear burning in the primary star in addition
to enhanced accretion luminosity and enhanced radiative trapping.

Stars with masses $\gtrsim 15\msun$ generate sufficient radiation
pressure to exceed their gravitational acceleration.  The second
column of figures \ref{f3}, \ref{f4} and \ref{f5} show that by then
the spherically-averaged radiation force from the central protostar
exceeds the inward gravitation attraction acting on the dusty envelope
of infalling gas in the case of $\Sigma=10.0~\textnormal{g cm}^{-2}$
and in the case of $\Sigma=2.0~\textnormal{g cm}^{-2}$ without
outflows.  In the case without outflows this strong radiative force
drives the expansion of a bubble of circumstellar gas away from the
central source.  The early development of this bubble can be seen in
the lower left panel of figure \ref{f3} at $t=0.4
t_{\textsubscript{ff}}$ and the radial extent of the bubble grows to a
size scale comparable to that of the initial core by $t=0.8
t_{\textsubscript{ff}}$ as shown in figure \ref{f5}.  The radiation
bubble emerges from the initial core on the left side of the density
slices in figure \ref{f5}.  This is due to the drift of the primary
star away from the center of mass of the initial core with time.  The
radiation bubble emerges first from the thinnest side of the initial
cloud, relative to the position of the primary star.  Accretion onto
the primary star continues through the radiatively supported bubble
via Rayleigh-Taylor unstable modes \citep{KrumholzRTBubbles,Jacquet}
that develop dense, radiatively self-shielding spikes of infalling
gas.  The evolution of radiative bubbles in similar simulations
without winds and without initial turbulence are discussed in detail
by \cite{KrumholzRTBubbles}.  The sole difference between the
radiation bubbles presented here and those in the earlier work is that
the bubbles presented here are considerably less symmetrical about the
central source owing to the turbulent ambient environment.

In the case with winds, the regions where the net force is dominated
by the outward radiation force lie within the outflow cavity.  These
regions are dominated by outflow irrespective of the radiation force.
With protostellar outflow, in no case does radiation force exceed that
of gravity acting on the infalling core gas.  Consequently, no such
radiation supported bubbles form in any of the models with
protostellar winds.  As predicted in \cite{KrumholzOutflows}, the
cavities evacuated by protostellar outflows provide sufficient
focusing of the radiative flux in the poleward directions that
accretion continues through the regions of the infalling envelope onto
the disk that are not disrupted by the protostellar wind shocks, and
the infalling motion of this gas is not interrupted by the effects of
radiation pressure.

%
%

%
%

%
%

%
%

\subsection{Protostar Properties}\label{properties}
\begin{figure}[tpb]
\begin{center}
\includegraphics[clip=true,width=0.49\textwidth]{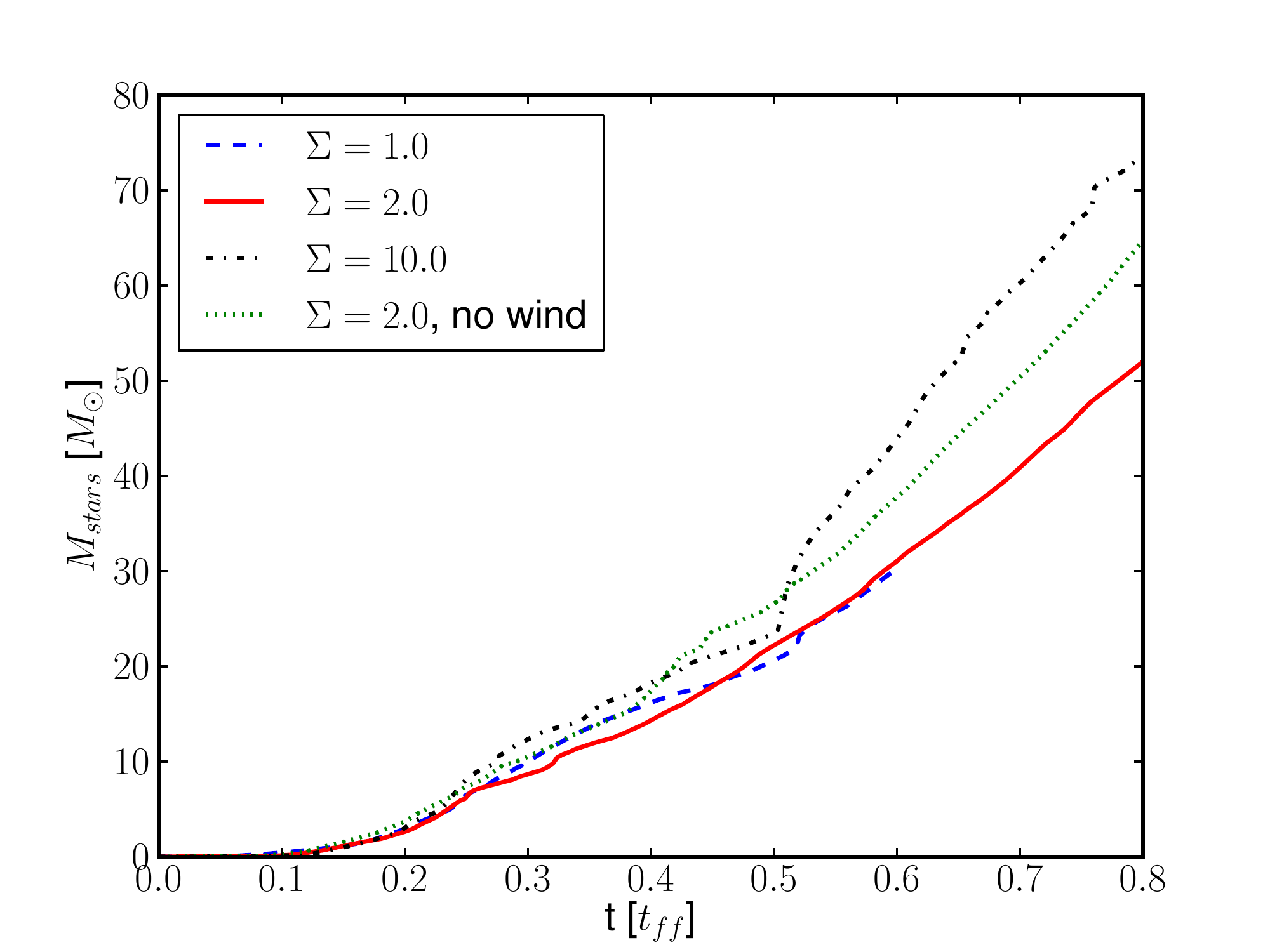}
\includegraphics[clip=true,width=0.49\textwidth]{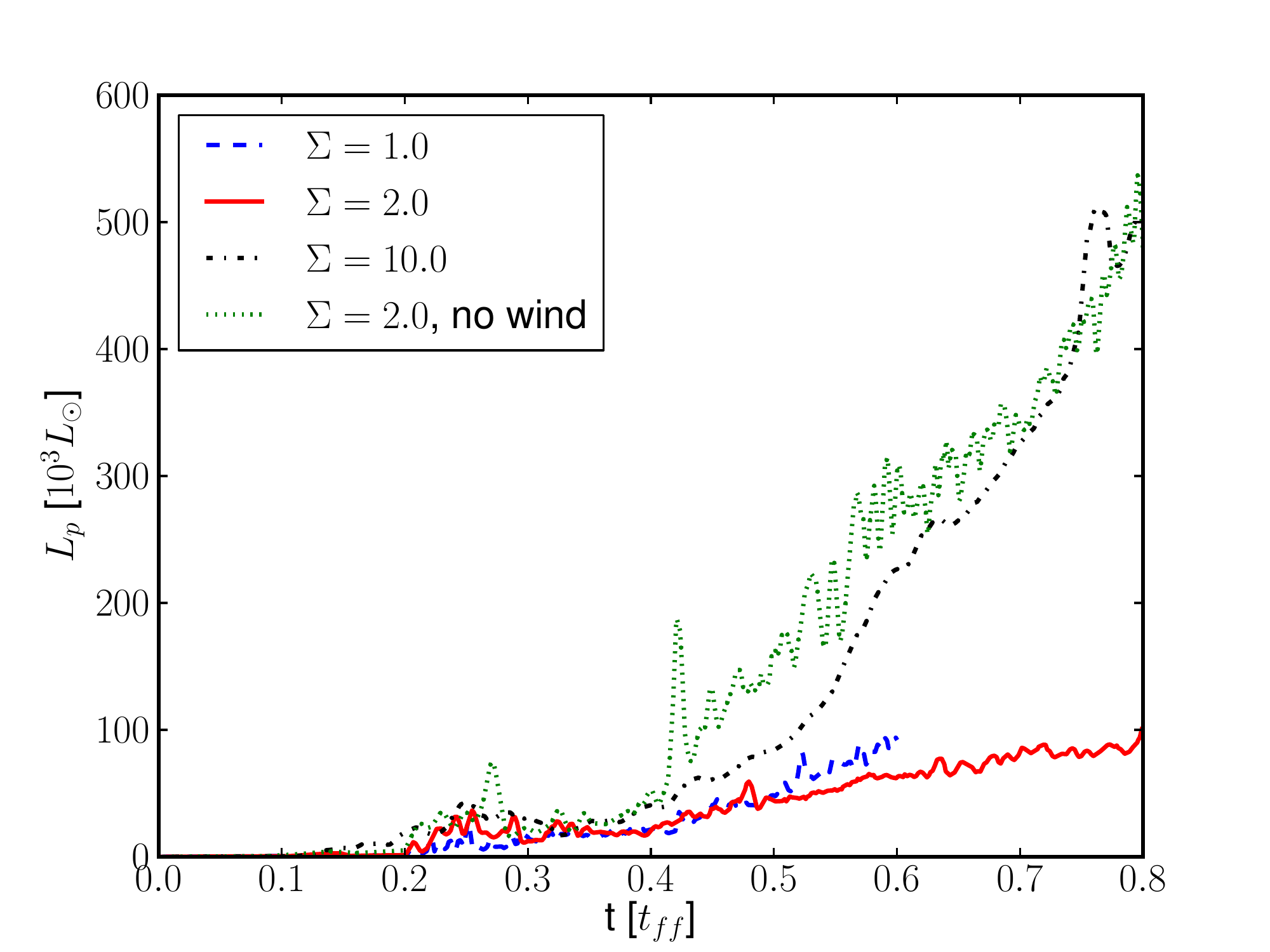} \\
\includegraphics[clip=true,width=0.49\textwidth]{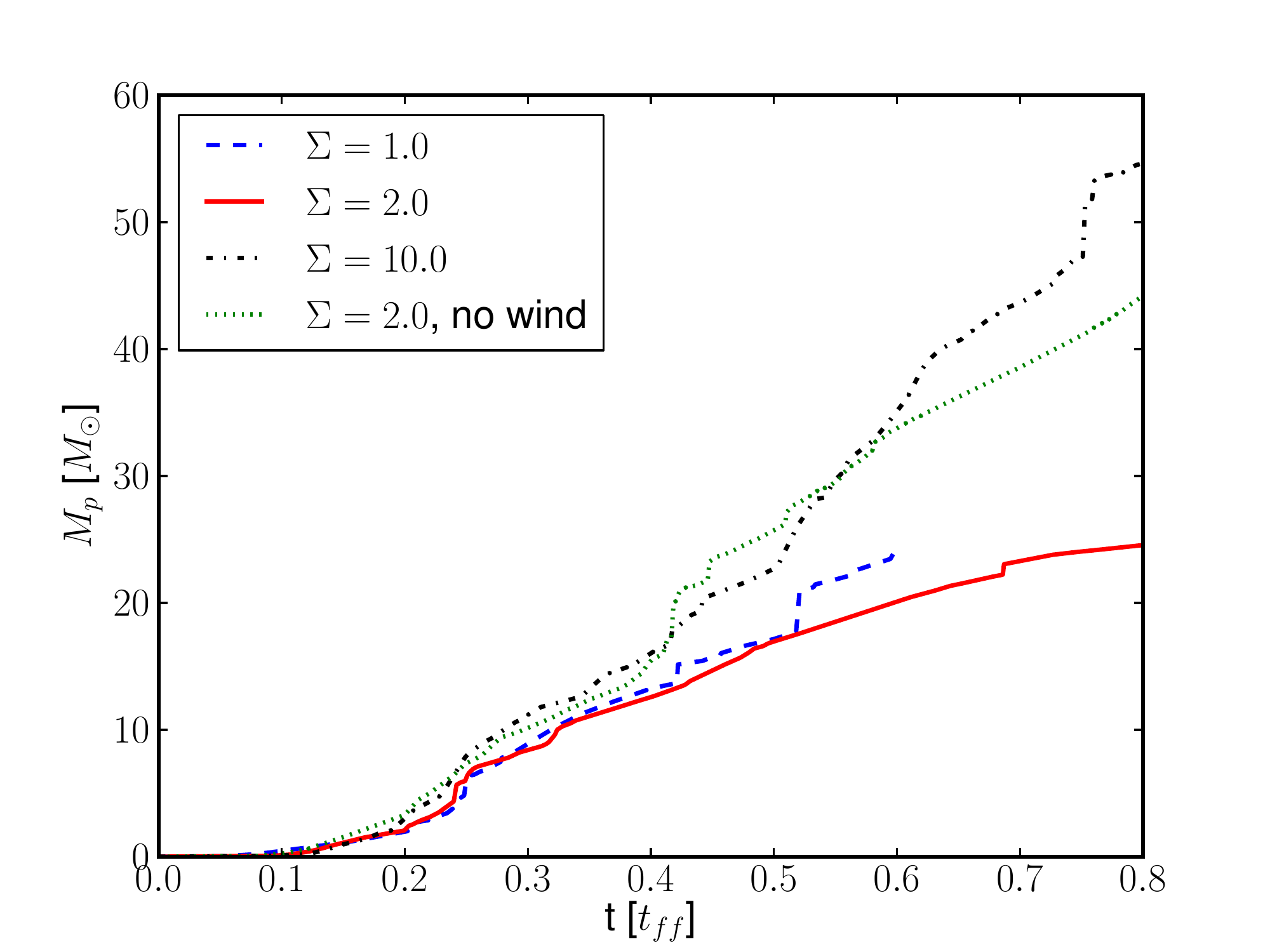}
\includegraphics[clip=true,width=0.49\textwidth]{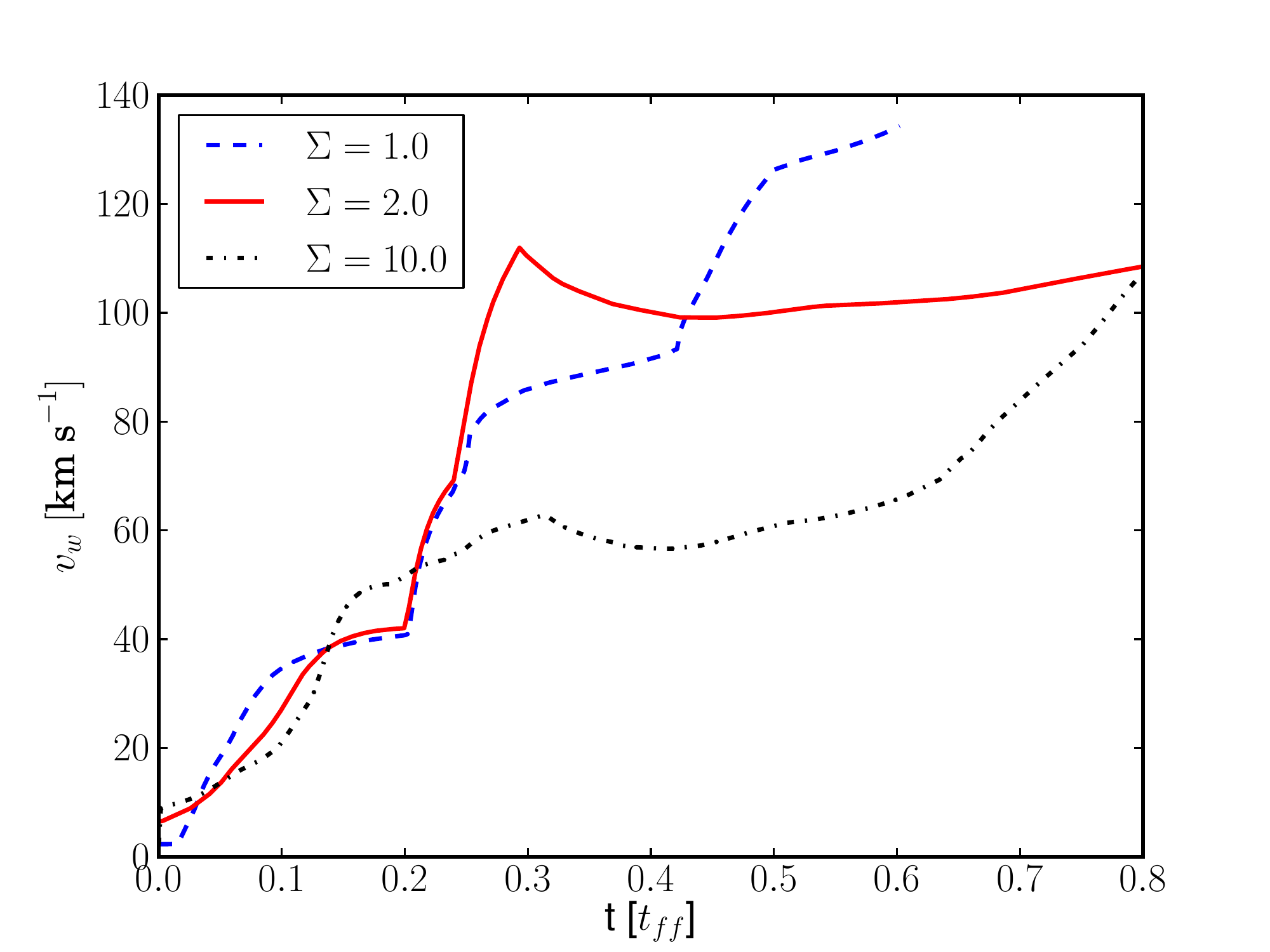} \\
\includegraphics[clip=true,width=0.49\textwidth]{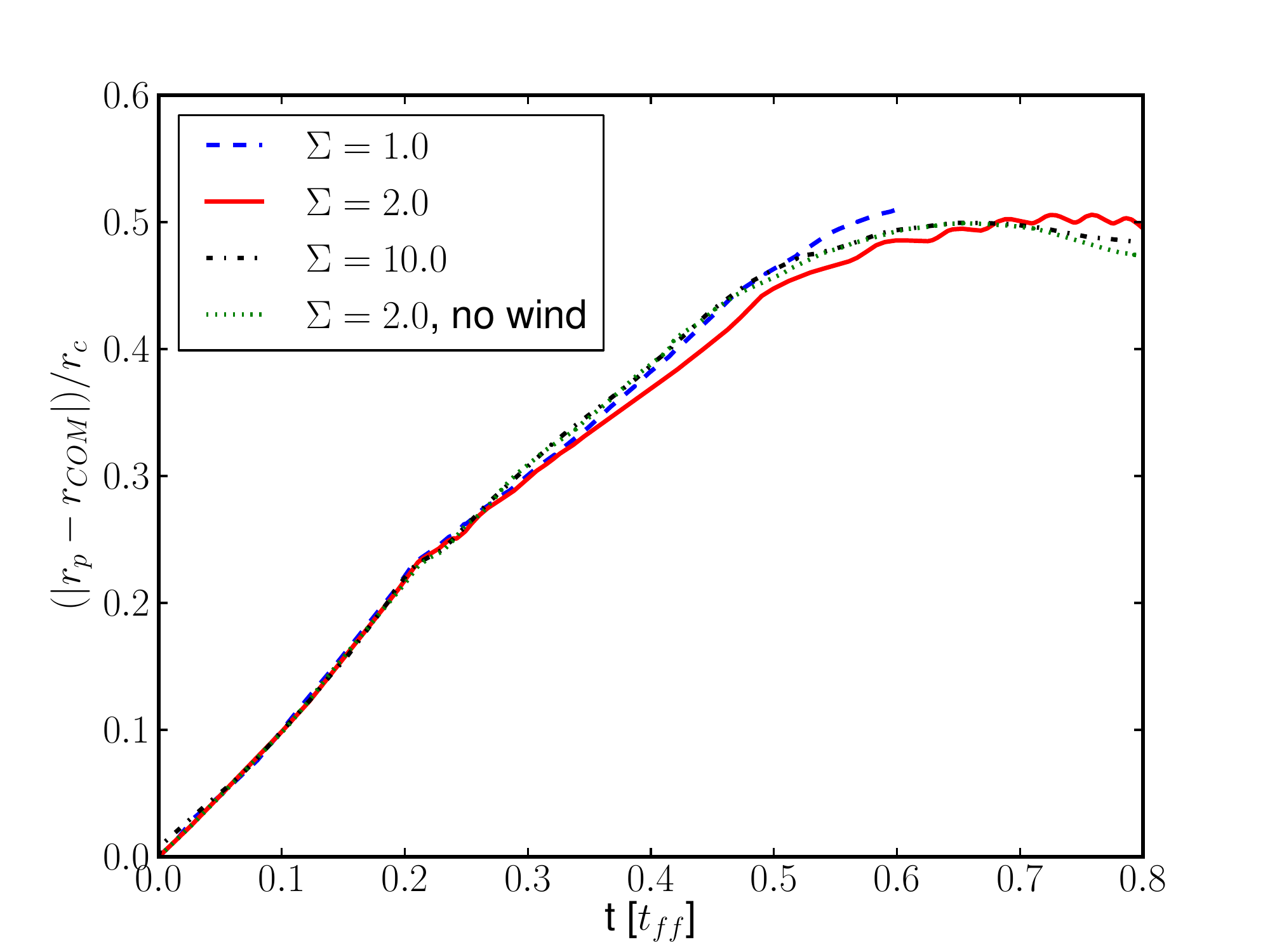}
\includegraphics[clip=true,width=0.49\textwidth]{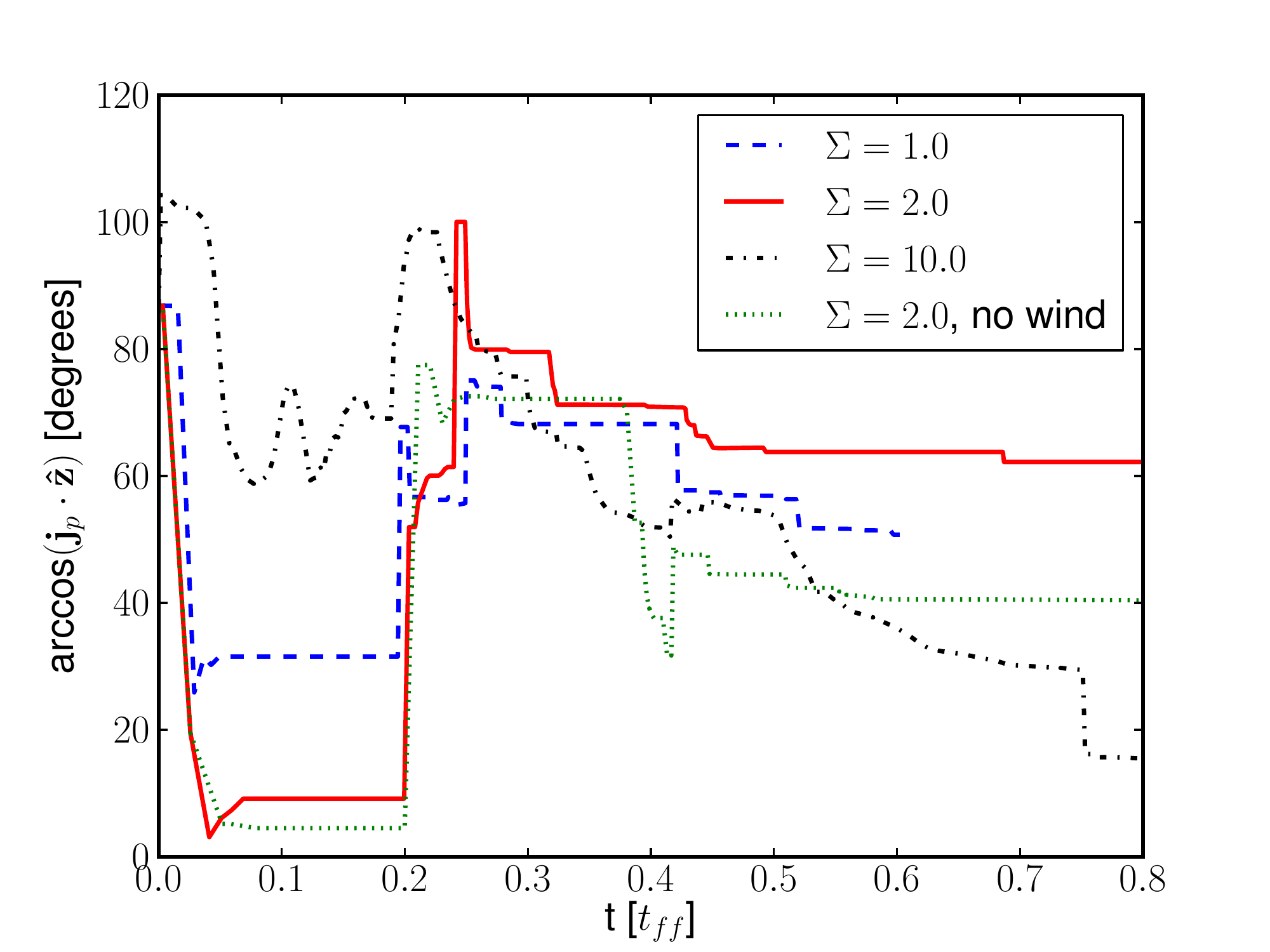} \\
\end{center}
\caption{Stellar properties as a function of time for each set of
  simulation parameters.  Upper Left: Total mass in stars. Upper
  Right: Primary star luminosity. The luminosity has been smoothed
  using a $200~\textnormal{yr}$ moving average to eliminate the high
  frequency contribution of the accretion luminosity in this plot.
  Middle Left: Primary star mass. Middle Right: Primary protostellar
  wind speed.  Lower Left: Position of the primary star relative to
  the center of mass of cloud.  Lower Right: Angle between the primary
  star's angular momentum vector and the z-axis.
  \label{f6}}
\end{figure}

The upper left panel of figure \ref{f6} shows the time dependence of
mass accretion onto star particles for each simulation and the middle
left panel shows the time dependence of mass accretion onto the
primary protostar.  Abrupt jumps in the primary protostellar mass
occur due to the merger of other particles that fall toward the
primary star.  The code has been constructed to merge star particles
that cannot be resolved on the resolution scale of four computational
zones on the finest level (see \cite{KrumholzSinks}).  Because these
mergers are resolution-dependent effects, we have taken care to assure
that the mass contribution to stellar sources due to mergers is small.
As we have discussed earlier, the ambient core in the case of
$\Sigma=1.0~\textnormal{g cm}^{-2}$ is subject to the least efficient
radiative heating and is the most susceptible to gravitational
fragmentation and therefore the most difficult to resolve.  The
particle mergers account for $\sim 15\%$ of the total primary
protostellar mass by the end of the simulation in the
$\Sigma=1.0~\textnormal{g cm}^{-2}$ case and about $\sim 10\%$ in the
higher surface density cases.  Most of the mass accumulated by merger
events in the case of $\Sigma=1.0~\textnormal{g cm}^{-2}$ occurs due to
the merger of a secondary star particle of $3.0 \msun$ at $t=0.52
t_{\textsubscript{ff}}$.  We do note, however, that \cite{myers} have found
that imposing a maximum mass threshold for mergers enhanced
fragmentation and limited the rate of mass growth of the primary star
in more highly resolved but otherwise similar simulations.  We
therefore regard the stellar mass predictions in the models presented
here as an upper limit.  In comparing the cases with protostellar
outflows we note that the total system mass in stars, $M$, exhibits a
weak trend toward more rapid accretion with higher surface density
even when the time is scaled by the free-fall time.  In other words,
$\dot{M} t_{\textsubscript{ff}}$ increases weakly with the surface density of
the initial core, $\Sigma$.  This is because outflows entrain and
unbind less gas from the ambient core in cases with higher surface
density.  Therefore, the luminosity output from the more massive
protostars is enhanced in the higher surface density cases.  This
contributes to more effective heating and decreased fragmentation of
the ambient core in the higher surface density cases as noted in
\S\ref{smallscale}.  Consequently, the simulations at lower surface
density fragment into low multiple systems earlier and are
characterized by significantly reduced accretion to the primary star
in comparison to the models at higher surface density.

The upper right and middle right panels of figure \ref{f6} show the
stellar luminosity and the speed of the protostellar wind driving from
the primary protostar as a function of time.  Nuclear burning
dominates the radiative output from stars with mass exceeding $5
\msun$ and the primary protostar dominates the stellar feedback into
the core.  The rapid increase in mechanical feedback from winds in all
of the models is driven by the Kelvin-Helmholtz contraction of
protostar \citep{shu87}.  Stellar contraction causes the escape speed
at the protostellar surface to increase which in turn increases the
wind ejection speed with $v_w = v_\textsubscript{esc}/3$ (see \S\ref{wind}).
By $t = 0.2 t_\textsubscript{ff}$ all of the models have undergone sufficient
contraction to ignite deuterium burning in the stellar cores, leading
to a rapid increase in luminosity.  Rapid contraction of the stellar
surface continues as deuterium is exhausted in the prestellar core,
giving rise to a commensurate increase in wind speed from $t = 0.2
t_\textsubscript{ff}$ to $t = 0.3 t_\textsubscript{ff}$.

Contrary to the increase in the effects of radiative feedback with
surface density, we find that the effects of mechanical feedback on
the ambient cores decrease with the initial surface density.  For $t >
0.4 t_\textsubscript{ff}$ the primary wind speeds show a noticeable decrease with
increasing surface density.  This occurs due to a mismatch between the
Kelvin time for the contraction of the stellar surface and the
free-fall time of the ambient molecular cloud core.  The former
depends mostly on the protostellar mass whereas the latter scales as
$t_\textsubscript{ff} \propto \Sigma^{-3/4}$ (equation (\ref{tff})).  This means that
lower surface density simulations form stars that undergo more rapid
Kelvin contraction {\it per unit free-fall time} and thereby eject
more powerful winds at equivalent stages of collapse.  We note that
this result is driven by the assumption built into our numerical model
that the wind ejection speed is proportional to the escape speed at
the protostellar surface.  As discussed in \S\ref{smallscale}, the
more powerful winds contribute to the enhanced disruption of the
ambient core in the lower surface density simulations.  However, the
overarching trend toward enhanced disruption of the ambient core with
lower surface density would remain even if the wind speed as a
function of core free-fall time were independent of surface density.
First, the outflow break-out time from the core, $t_\textsubscript{outflow} =
r_c/v_w$ would scale, relative to the free-fall time, as $t_\textsubscript{
outflow}/t_\textsubscript{ff} \sim \Sigma^{1/4}$, resulting in later outflow
emergence from higher surface density cores.  Second, higher surface
density cores enhance the confinement of outflow cavities due to
higher ambient pressures.

The cascade of turbulent motions through the collapsing cloud strongly
affects the motion of the primary star as a function of time.  The
lower left panel of figure \ref{f6} plots the distance between the
primary star, at position $\vec{r_p}$, and the center of mass of the
system, at position $\vec{r}_{COM}$.  The drift of the primary star
away from the center of the system is a result of the random velocity
field.  Because most of the turbulent energy is in large wavelength
modes, the denser center of the cloud will tend to have a different
velocity than the lower density edges.  The star initially forms
in a compressional mode that is much larger than the local reservoir
of gas that starts the protostar, but smaller than the diameter of the
core.  This initial reservoir of gas that starts the star is not
co-moving with the center of mass of the core.  Therefore, while the
star does initially form at the bottom of the gravitational potential
well it has a finite kinetic energy and can oscillate within the
well.  As a result the location of the star drifts from where it
first forms in our simulations.

The cascade of turbulent motions through the collapsing cloud also
strongly affect the orientation of the primary star as a function of
time.  The lower right panel of figure \ref{f6} shows the angle of
inclination of the angular momentum accreted by the primary protostar
particle relative to the $\hat{\vec{z}}$ direction.  We note that the
total angular momentum of the initial cloud has an orientation that is
inclined $27^\circ$ to the $\hat{\vec{z}}$ direction. Because no radiative
heating feedback is present in the initial state of the cloud, the
early evolution of the simulations are characterized by fragmentation
of the densest portion of the initial cloud into 2 to 4
gravitationally bound particles, as shown in figure \ref{f2}.  By
$t=0.3 t_{\textsubscript{ff}}$, the radiative feedback from the primary star
is sufficient to suppress local fragmentation and the initial
fragments merge.  The angular momentum of the primary particle at
early time varies over $\sim 90^\circ$ as a consequence of variation
in the angular momentum of the surrounding gas and as a consequence of
the coalescence of the initial fragments.  After $t=0.3
t_{\textsubscript{ff}}$ the primary star has built up a sufficient moment of
inertia relative to the rate of angular momentum deposition by
accretion that the orientation of the star changes less rapidly.
Subsequent evolution ($t=0.3 t_{\textsubscript{ff}}$ to $t=0.8
t_{\textsubscript{ff}}$ ) of the orientation of the primary star is
characterized by a rotation of $20^\circ$ to $40^\circ$.  We note that
the change in angular momentum in the lower surface density cases
($\Sigma=1.0~\textnormal{g cm}^{-2}$ and $\Sigma=2.0~\textnormal{g cm}^{-2}$)
occurs in abrupt jumps, whereas the higher surface density case is
less susceptible to numerical fragmentation and therefore are
characterized by relatively smoother change.

%

\begin{figure}[tpb]
\begin{center}
\includegraphics[clip=true,width=0.49\textwidth]{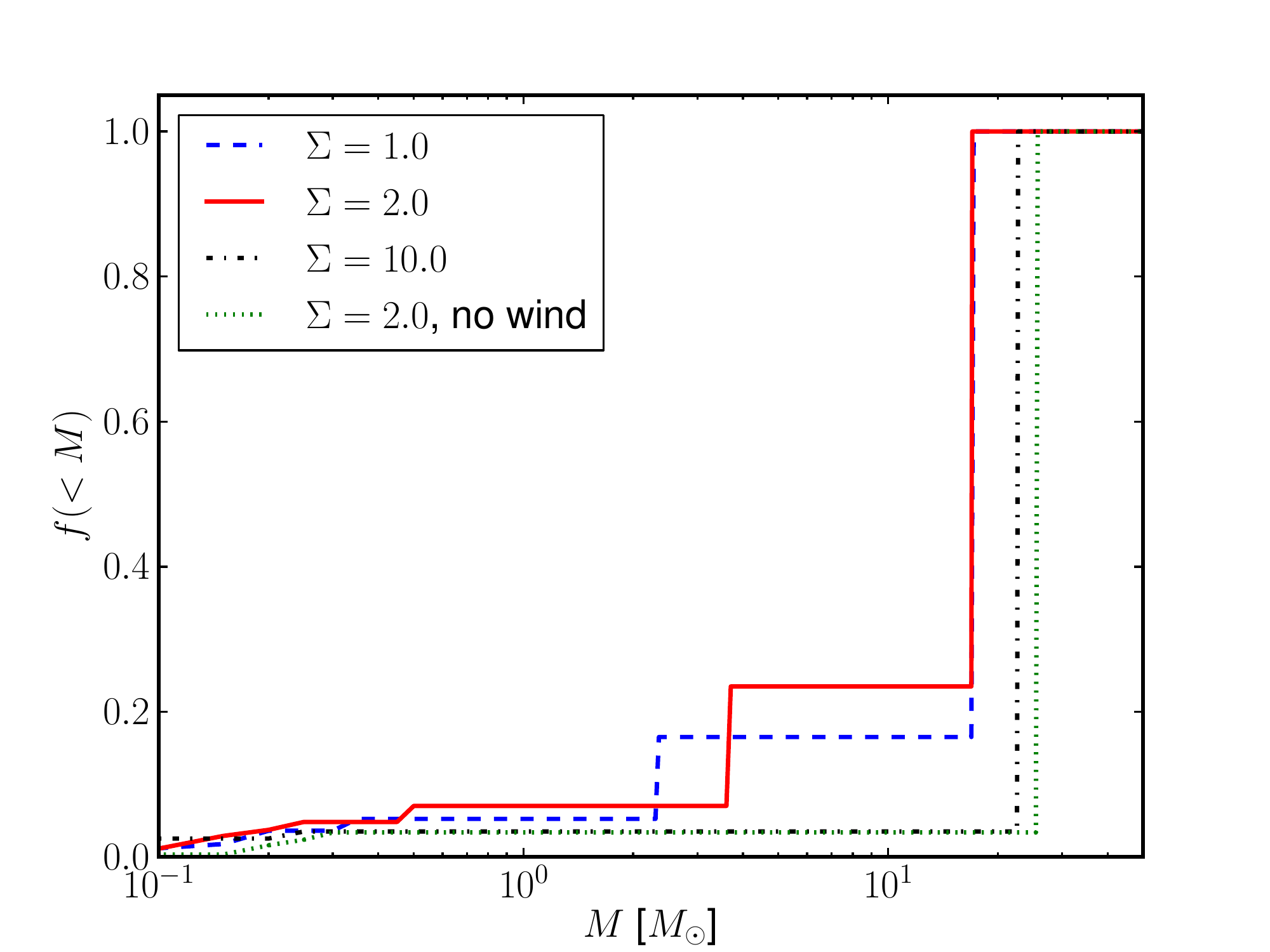}
\end{center}
\caption{ Fraction $f(<M)$ of total stellar mass contained in stars
  with mass $<M$ as a function of the total mass in stars for each of
  the runs at time $t=0.5 t_\textsubscript{ff}$.
\label{f7}}
\end{figure}

The cumulative distribution of stellar masses at $t=0.5 t_\textsubscript{ff}$
for each of the runs shown in figure \ref{f7}.  Both the highest
surface density case with $\Sigma = 10.0~\textnormal{g cm}^{-2}$ and the
case without outflows at surface density $\Sigma = 2~\textnormal{g
  cm}^{-2}$ collapse to a single star of $70\%$ of the total mass
accreted, consistent with the qualitative similarity in the
small-scale temperature structure noted in \S\ref{smallscale}.  The
lower surface density cases on the other hand fragment into binary
systems with $> 1~\msun$ secondaries present by $t=0.4 t_\textsubscript{ff}$.
These results demonstrate that the absence of outflows and/or higher
initial surface densities result in less fragmentation and the
production of fewer, more massive stars.

%
%

\subsection{Radiation Focusing}\label{focusing}
The radiative feedback from stellar sources into the ambient cloud is
not isotropic.  Four possible causes of anisotropic stellar output
include (1) the clearing of an optically thin outflow-swept cavity
along the poles of the star, (2) shielding in the midplane due to the
presence of an optically thick accretion disk, (3) non-uniform optical
thickness of the ambient cloud due to the displacement of the star
relative to the center of mass of the ambient cloud, and (4)
non-uniform optical thickness of the ambient cloud due to the
turbulent density structure.  In this section we consider the
distribution of radiative output from the primary star in each
simulation as a function of solid angle to elucidate the importance of
each of these effects in shaping the radiative feedback into the
ambient cloud.

\begin{figure}[tpb]
\begin{center}
\includegraphics[clip=true,width=1.\textwidth]{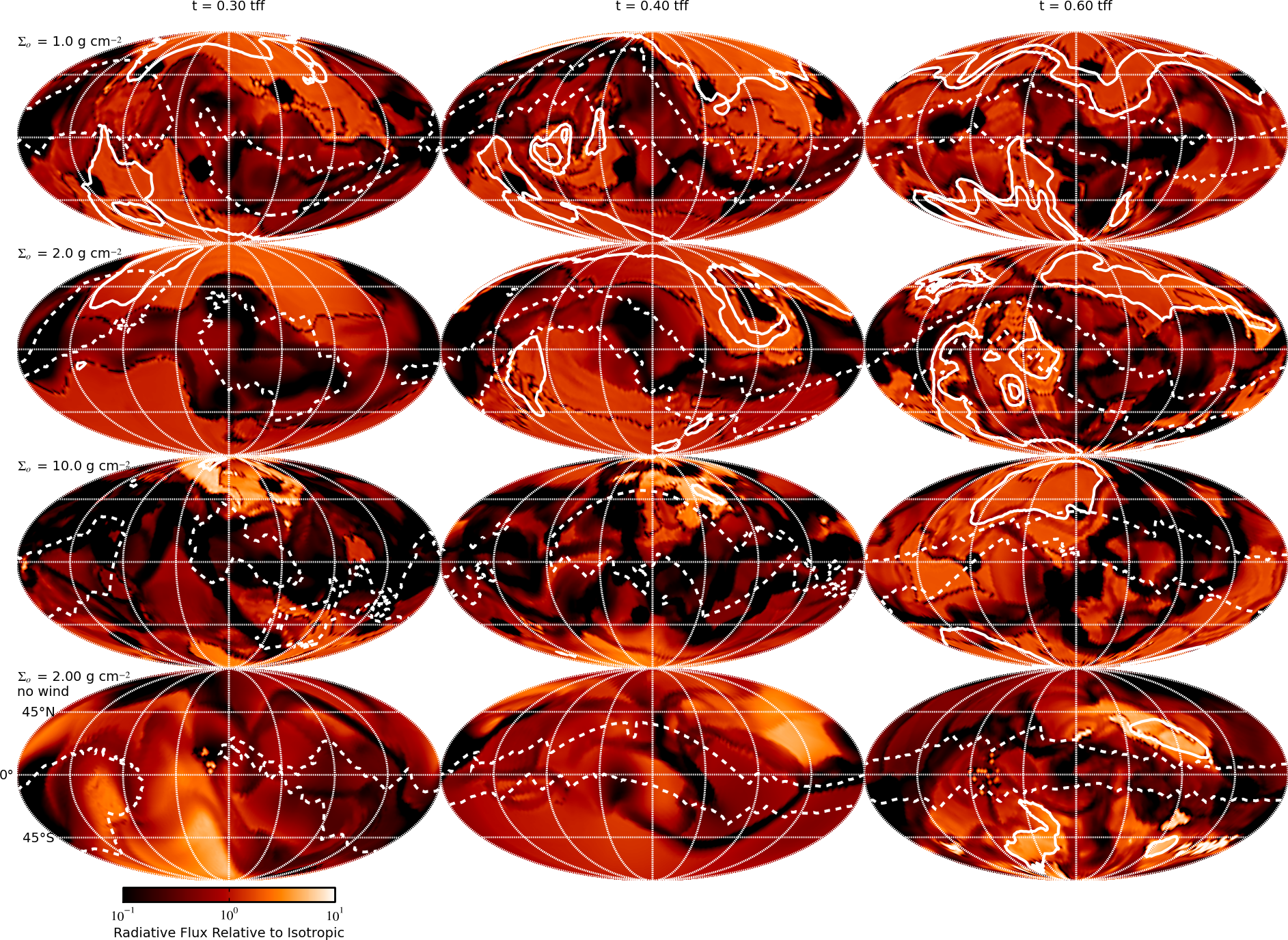}
\end{center}
\caption{Radial component of the radiation flux, normalized to the
  isotropic flux at $1500~\textnormal{AU}$ from the primary
  protostar. Columns indicate the times $t=0.3t_\textsubscript{ff}$,
  $t=0.4t_\textsubscript{ff}$ and $t=0.6t_\textsubscript{ff}$ from
  left to right and the rows indicate the simulation parameters of
  $\Sigma = 1.0~\textnormal{g cm}^{-2}$, $\Sigma = 2.0~\textnormal{g
  cm}^{-2}$, $\Sigma = 10.0~\textnormal{g cm}^{-2}$ and $\Sigma =
  2.0~\textnormal{g cm}^{-2}$ without winds from top to bottom.  The
  coordinate system is defined such that the angular momentum of the
  primary star points northward and the azimuthal coordinate facing
  the thinnest edge of the cloud due to the motion of the star
  relative to the center of mass of the system is centered in each of
  the plots.  Contours of the $75^{\textnormal{th}}$ percentile column
  density from $r=0~\textnormal{to}~1500~\textnormal{AU}$ are shown as
  dashed lines and contours of the radial velocities of
  $20~\textnormal{km s}^{-1}$ and $50~\textnormal{km s}^{-1}$ at
  $r=1500~\textnormal{AU}$ are shown as solid lines.
\label{f8}}
\end{figure}

Figure \ref{f8} shows the radiative flux from the primary protostar as
a function of spherical angle in each model at times $t=
0.3,~0.4,~\textnormal{and}~0.6 t_\textsubscript{ff}$, normalized to the radiative
flux that would be expected in an isotropic environment $\vec{F}_\textsubscript{
isotropic} \cdot \hat{\vec{r}}= L_p/(4 \pi r^2)$.  The plots shown in
figure \ref{f8} have been constructed from slices of the radial
radiation flux at a distance of $1500~\textnormal{AU}$ from the primary
star, rotated into the coordinate system where the angular momentum of
the primary star points upward.  The angular distribution of the
radiative flux is relatively insensitive to the position of the
spherical slice, provided that the slice has radius greater than that
of the disk around the primary protostar.  In varying the radius of
the spherical slice from $1500~\textnormal{AU}$ to $10^4~\textnormal{AU}$, we
note a decreased contrast between regions of low and high radial flux
by about $\sim 25\%$, which we attribute, in part, to the flux limited
diffusion approximation used for the radiation transport.  However, the
location and extent in solid angle of the large-scale features is
independent of the position of the slice. The azimuthal coordinate
facing the thinnest edge of the cloud due to the motion of the star
relative to the center of mass of the system is centered in each of
the plots.  Regions of peak outward radiative flux correlate very well
with outflow ejection, and regions of low outward radiative flux
correlate very well with the presence of the accretion disk near the
midplane.  The clearing of low optical depth paths of escape by
outflow ejection and shielding by the dense accretion disk in the
midplane are therefore the dominant effects in focusing the radiative
feedback from the star.  As gas falls in toward the primary protostar,
it carries its angular momentum with it. In the highest surface
density gas the accretion flow is fairly smooth because radiative
heating raises the pressure near the primary star and prevents gas
around it from clumping up. At lower surface densities, however, the
accreting gas may be partially collapsed under its own gravity, or may
even have collapsed completely to form stars that then merge with the
primary. As a result, angular momentum tends to be accreted in
distinct lumps, leading to rapid reorientation of the accreting star
over short timescales.  The radiative flux is far more focused as
bipolar in the case of $\Sigma=10.0~\textnormal{g cm}^{-1}$, consistent
with the narrower geometry of the outflow cavity in this case.

\begin{figure}[tpb]
\begin{center}
\includegraphics[clip=true,width=0.5\textwidth]{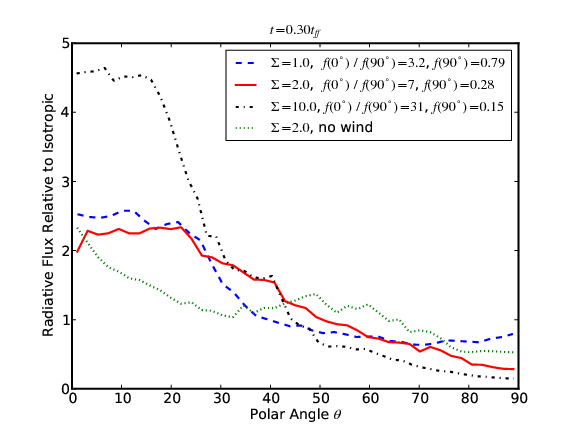} \\
\includegraphics[clip=true,width=0.5\textwidth]{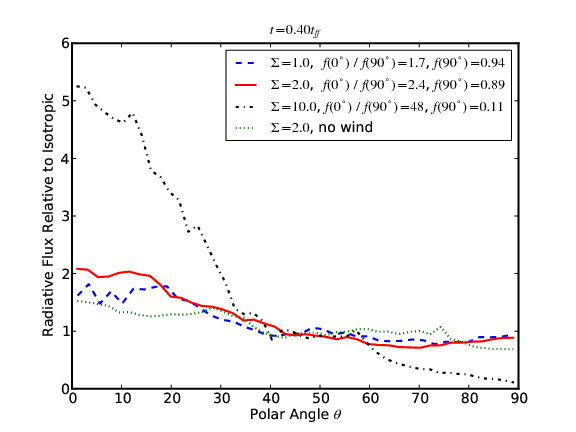} \\
\includegraphics[clip=true,width=0.5\textwidth]{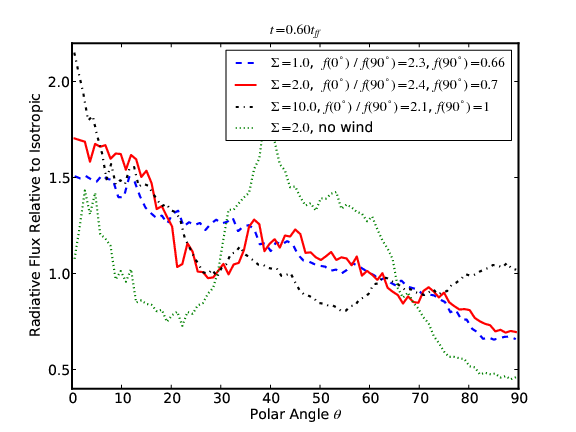}
\end{center}
\caption{Radial component of the radiation flux at
$1500~\textnormal{AU}$ in radius from the primary star, normalized to
the isotropic flux, as a function of polar angle. The flux shown at a
given $\theta$ is a volume average over a pair of rings at polar
angles $\theta=0$ and $180^\circ-\theta$ that cover all azimuthal
angles $\phi$.  The coordinate system is oriented so that $\theta=0$
corresponds to the rotation axis of the primary star and the direction
in which the wind is launched.
\label{f9}}
\end{figure}

Figure \ref{f9} shows the azimuthally-averaged radiative flux from the
primary protostar as a function of polar angle in each model at times
$t= 0.3,~0.4,~\textnormal{and}~0.6 t_\textsubscript{ff}$.  At each of these times,
the effect of the presence of protostellar outflow cavities is clearly
evident showing polar radiation flux $1.7$ to $15$ times that at the
midplane.  The models with protostellar outflow show enhancement of
the polar flux relative to the case without outflows by comparable
factors.  The degree of poleward focusing of the radiation flux
diminishes with time due to broadening of the outflow evacuated
cavities that focus the radiation.  We note that polar flux in the
$\Sigma = 1.0~\textnormal{g cm}^{-2}$ and $\Sigma = 2.0~\textnormal{g
cm}^{-2}$ cases at $t=0.4 t_{\textsubscript{ff}}$ underrepresents the flux
focusing due to outflow cavities because the outflow cavities are
tilted by $\sim 20^\circ$ relative to the orientation of the primary
protostar shown in figure \ref{f8}.  Similar radiation focusing
effects were also shown in \cite{KrumholzOutflows}, where the authors
considered the effects of the presence of outflow cavities on
radiation escape from the infalling envelope around massive
protostars.  Using static radiative transfer models they showed that
focusing of the radiative force from the central star throughout the
outflow cavity results in a reduction of the equatorial radiative flux
relative to a control model without outflows by factors of $1.7$ to
$14$, depending on the width and shape of the region of low optical
depth in the outflow cavity.  Therefore, the models presented here
support the \cite{KrumholzOutflows} prediction that outflows reduce
the Eddington radiation pressure barrier to high-mass star formation
by reducing the radiation force exerted in the infalling cloud gas.
However, it has also been shown that fully 3D Rayleigh-Taylor modes
can remove the Eddington barrier even when protostellar outflows are
neglected \citep{KrumholzRTBubbles}.

%
%

%
%

\subsection{Star Formation Efficiency} \label{sfe}
\cite{MatznerCores} give an analytic model for the core to star
formation efficiency 
\begin{equation}
\epsilon_\textsubscript{core}=1-M_\textsubscript{ej}/M,
\end{equation}
where $M_\textsubscript{ej}$ is the net mass ejected from the core by
entrainment into the protostellar outflow and $M$ is the initial core
mass.  For the case of an unmagnetized core the model predicts
\begin{equation}
\epsilon_\textsubscript{core} = \frac{2 X(c_g)}{1+\sqrt{1+4(1+f_w)^2 X^2(c_g)}} \label{prediction},
\end{equation}
where the dimensionless function $X$ is defined by
\begin{equation}
X  = c_g \ln\left(\frac{2}{\theta_0}\right) \frac{v_\textsubscript{esc}}{f_w \bar{v}_w},
\end{equation}
$v_\textsubscript{esc}$ is the escape speed from the edge of the
core, and the parameter $c_g$ depends on the core density profile and
free-fall time relative to the effective timescale for the wind
driving.  For an unmagnetized core with profile $\rho \propto
r^{k_\rho}$, equation (A19) of \cite{MatznerCores} provides an estimate
of 
\begin{equation}
c_{g,\gg 1}=\frac{\pi(9-3k_\rho)(4-k_\rho)}{8-3k_\rho}\left(\frac{t_w}{t_\textsubscript{ff}}\right)
\end{equation}
for steady winds.  The estimate depends on the the age of the steady
wind, $t_w$ and is valid for $c_g \gg 1$.  In the limit of an
impulsively driven wind, $t_w \rightarrow 0$, \cite{MatznerCores}
provide an estimate of
\begin{equation}
c_{g,0}=\sqrt{\frac{9-3k_\rho}{8-3k_\rho}}.
\end{equation}
To compare our simulations with the analytic model, we choose $c_g$ by
interpolating between the two limits as:
\begin{equation}
c_g=\frac{\pi(9-3k_\rho)(4-k_\rho)}{8-3k_\rho}\left(\frac{t_w}{t_\textsubscript{ff}}\right)+\sqrt{\frac{9-3k_\rho}{8-3k_\rho}}.
\end{equation}
For $k_\rho = 3/2$ this expression becomes
\begin{equation}
c_g=2.52\frac{t_w}{t_\textsubscript{ff}}+1.13.
\end{equation}
The mass-weighted average wind speed that characterizes the wind
momentum injection into the core is
\begin{equation}
\bar{v}_w = \sum_\textsubscript{stars} \frac{1}{f_w M_i} \int \dot{M}_{w,i} v_{w,i} dt. \label{vbar}
\end{equation}
Values of $\bar{v}_w$ and $v_{\textsubscript{esc}}$ are given in table
\ref{t2}.  Because our numerical wind injection approach is based on
volume-averaged quantities inside of an 8-cell radius wind injection
sphere as described in \S\ref{wind}, the effective numerical
flattening parameter is $\theta_{0,{\rm eff}} = 5.75 \times 10^{-4}$
for winds with an opening angle $> 32^\circ$ and we shall use this as
the flattening parameter for the purposes of comparing the simulations
to the analytic model.  We note that $X$ depends logarithmically on
the flattening parameter (equation (\ref{prediction})) and therefore
the model prediction is not very sensitive to the estimate of the
effective numerical flattening angle.

\begin{figure}[tpb]
\begin{center}
\includegraphics[clip=true,width=0.49\textwidth]{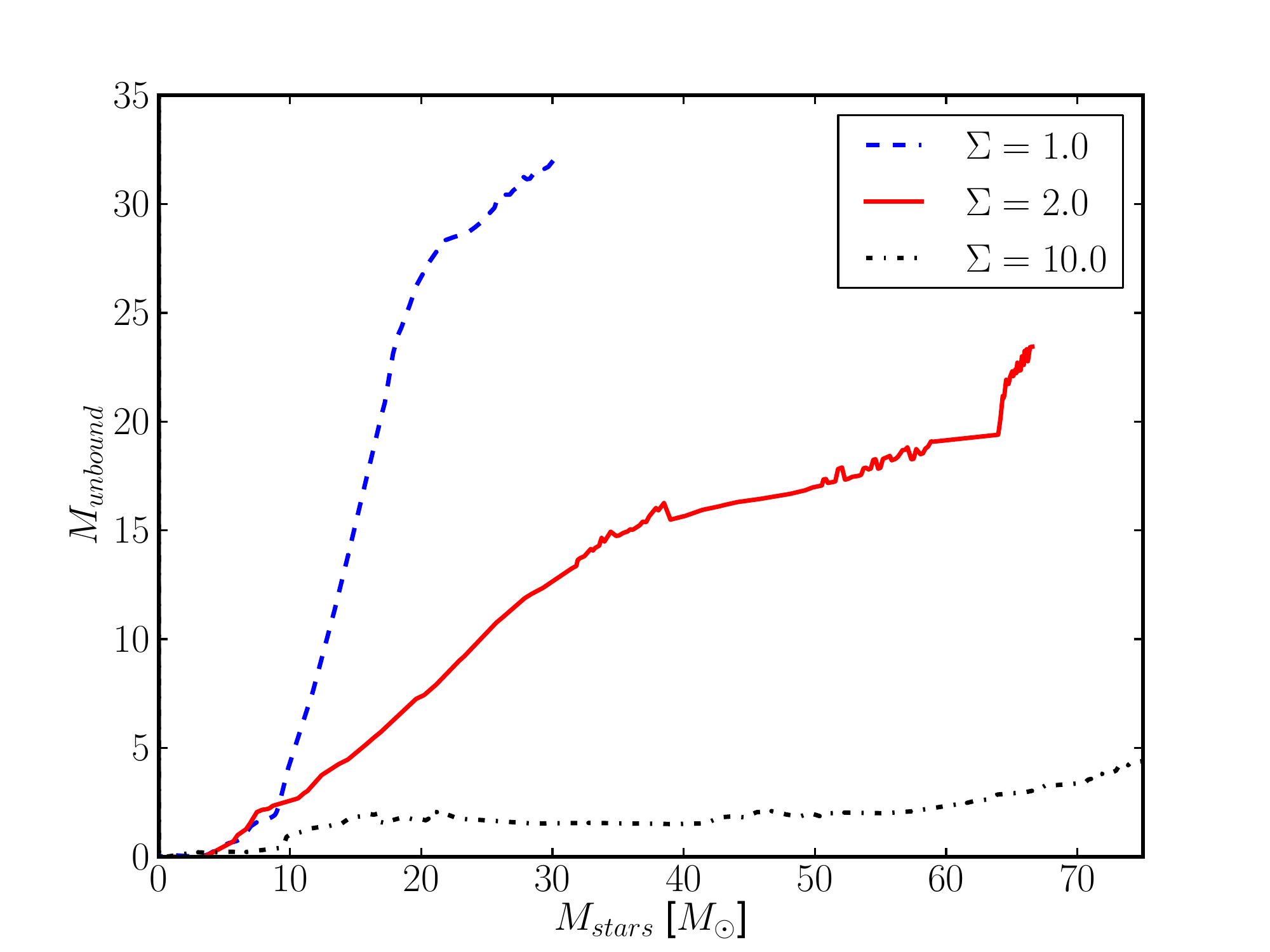}
\includegraphics[clip=true,width=0.49\textwidth]{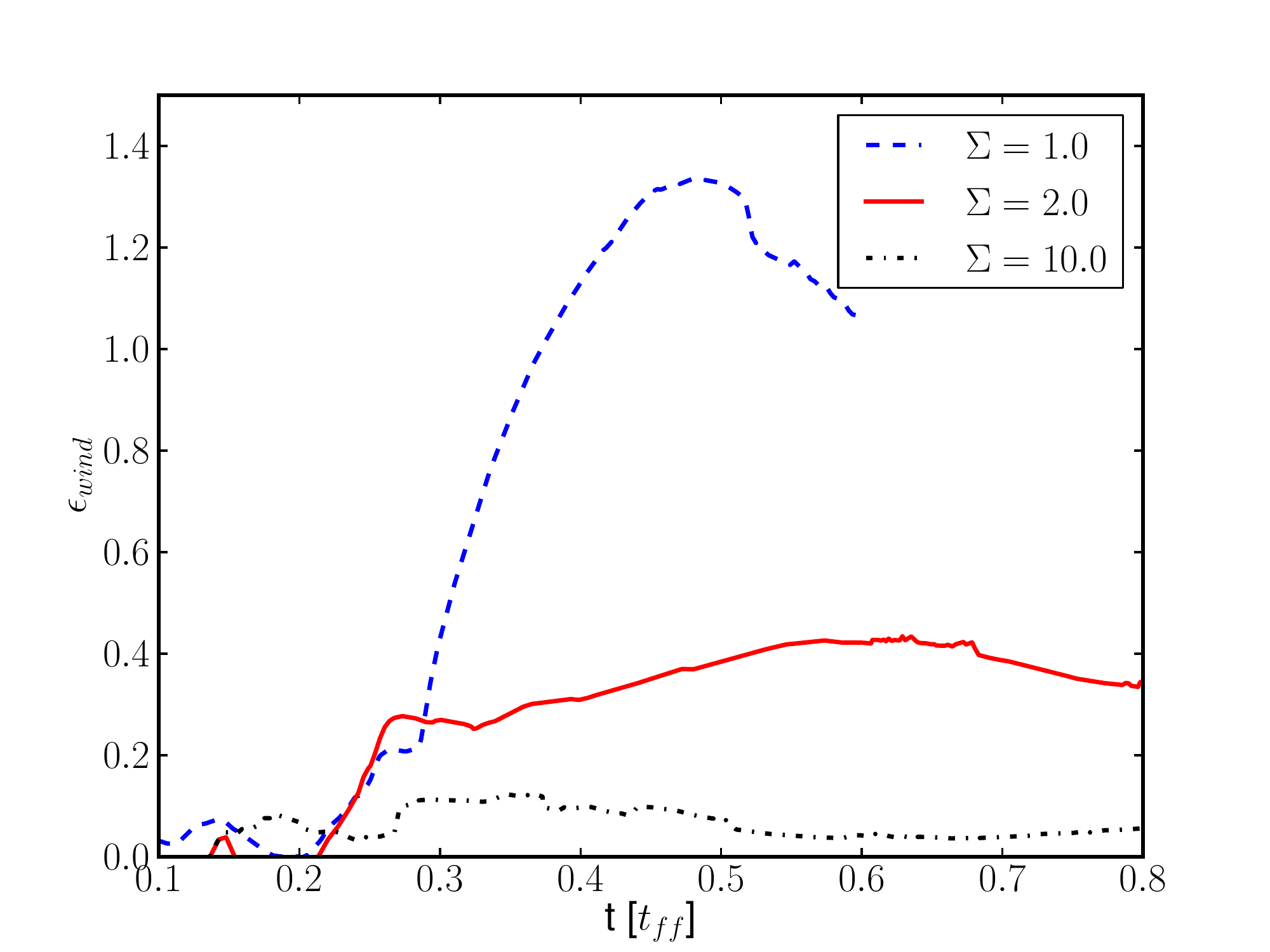}
\end{center}
\caption{Left: Mass with outward radial speed greater than the escape
speed of the system as a function of the total mass in stars. Right:
Mass with outward radial speed greater than the escape speed relative
to the total mass in stars as a function of time. \label{f10}}
\end{figure}

Due to constraints on computational time, we have not run numerical
simulations sufficiently long to determine the final star formation
efficiency.  To facilitate comparison between the numerical and
analytic model, we focus our attention to the ratio of the mass ejected
by winds to the total stellar mass,
\begin{equation}
\epsilon_\textsubscript{wind} = \frac{M_\textsubscript{ej}}{\sum_\textsubscript{stars} M_i} = \frac{1-\epsilon_\textsubscript{core}}{\epsilon_\textsubscript{core}}, \label{windprediction}
\end{equation}
where $M_\textsubscript{ej}$ is the total mass ejected from the
system.  This quantity can be computed as a function of time
throughout the simulation.  For the purpose of comparing our numerical
simulations to the analytic prediction, we heuristically define the
ejected mass as any mass that has been either ejected from the
simulation domain or that is propagating with a sufficient radial
component of velocity away from the center of mass of the system to
overcome its gravitational binding to the system.  The left panel of
figure \ref{f10} shows a plot of the total wind mass ejected in each
simulation as a function of the total mass in stars and the right
panel shows the simulation result for the wind ejection efficiency
$\epsilon_\textsubscript{wind}=M_\textsubscript{ej} /
\sum_\textsubscript{stars} M_i$ as a function of time for each
simulation.  We note that wind-launched gas in the highest surface
density simulation with $\Sigma=10.0~\textnormal{g cm}^{-2}$ emerges
from the initial core relatively late in the simulation at $t \sim 0.6
t_\textsubscript{ff}$.  At the end of the simulation, much of the
wind-launched gas is still entrained in the limbs of the outflow
cavity.  This is a transient effect that is not included in the
analytic model and we therefore will focus our attention in comparing
the analytic prediction to only the $\Sigma = 1.0~\textnormal{g
cm}^{-2}$ and $\Sigma = 2.0~\textnormal{g cm}^{-2}$ models.  By
inspection of figure \ref{f10}, we adopt the values of $t_w = 0.5
t_\textsubscript{ff}$ and $t_w = 0.3 t_\textsubscript{ff}$ as
characteristic of the age of the winds in the $\Sigma =
1.0~\textnormal{g cm}^{-2}$ and $\Sigma = 2.0~\textnormal{g cm}^{-2}$
simulations respectively.

The analytic predictions given by equations (\ref{prediction})
\& (\ref{windprediction}) for the outflow mass-weighted wind speed
$\bar{v}_w$ for each simulation are given in table \ref{t2}.  The wind
ejection efficiency at the end of each simulation $(t =
t_{\textsubscript{end}})$ are listed in the table as
$\epsilon_\textsubscript{wind,~simulation}$ for comparison to the model
predictions.

%
%
\begin{deluxetable}{l r r r r}
  \tablewidth{0pt}
  \tablecaption{Outflow ejection.}\label{t2}
  \startdata
  \tableline
  $\Sigma~(\textnormal{g cm}^{-2})$ & 1.0 & 2.0 & 10.0 \\
  $t_\textsubscript{end}~(t_{\textsubscript{ff}})$ & 0.6 & 0.8 & 0.8 \\
  $v_\textsubscript{esc}~(\textnormal{km s}^{-1})$ & 4.27 & 5.08 & 7.60 \\
  $\bar{v}_w|_{t=t_\textsubscript{end}}~(\textnormal{km s}^{-1})$ & 87.7 & 72.2 & 71.0 \\
  $\epsilon_\textsubscript{core}$ & 0.70 & 0.73 & \nodata \\
  $\epsilon_\textsubscript{wind,~simulation}$ & 1.06 & 0.342 & 0.0563 \\
  $\epsilon_\textsubscript{wind}$ & 0.42 & 0.370 & \nodata \\
  \enddata
  \tablewidth{\textwidth}
  \tablecomments{Simulation results (rows 1-5) and analytic
model predictions (rows 6 \& 7).  The columns indicate the cases of
$\Sigma = 1.0~\textnormal{g cm}^{-2}$, $\Sigma = 2.0~\textnormal{g
cm}^{-2}$ and $\Sigma = 10.0~\textnormal{g cm}^{-2}$ from left to
right.  As discussed in the text, the $\Sigma = 10.0~\textnormal{g
cm}^{-2}$ simulation was not evolved sufficiently far in time to
compare with the analytic model.}
\end{deluxetable}

In drawing conclusions from comparing the analytic model result for
the wind ejection ratio $\epsilon_\textsubscript{wind}$ to the simulation
result $\epsilon_\textsubscript{wind,~simulation}$ it is important to bear in
mind that the analytic model predicts the fraction of the initial core
that is ejected by a wind over the entire course of evolution of the
initial core, whereas the numerical model is only averaged over the
duration of the core evolution up to the end of the simulation.
Furthermore, the analytic model cannot account for the time dependence
associated with the propagation of protostellar wind shocks through
the highly inhomogeneous ambient core.

For the lowest surface density case $\Sigma = 1.0~\textnormal{g
cm}^{-2}$, the simulation has ejected more mass with the wind per unit
of mass accretion than predicted by the analytic model by a factor of
$2.3$.  This is likely due to efficient entrainment of core gas into
the wind via the interaction of the protostellar winds as they
propagate through the network of dense filaments in the ambient core
at early time.  By $t=0.4~t_\textsubscript{ff}$, a wide wind cavity has
been cleared through the initial core in the simulation (see the top
row of figure \ref{f3}).  The time-integrated outflow ejection
behavior in the simulation is biased toward this early-time behavior
because the simulations are only advanced to $t=0.6
t_{\textsubscript{ff}}$ in the $\Sigma = 1.0~\textnormal{g cm}^{-2}$
case and $t=0.8 t_{\textsubscript{ff}}$ in the other models.  In
figure \ref{f10}, we note that evolution of the system past
$0.5t_{\textsubscript{ff}}$ carries forward with less efficient
entrainment of core gas into the outflows as the outflowing gas at
later time propagates unimpeded through the wind channel.  We expect
that with continued evolution the system would asymptote to a steady
state value of $\epsilon_\textsubscript{core}$ that is closer to the
model prediction.  However, constraints of computational cost have
prevented us from testing this expectation.

The intermediate surface density case $\Sigma = 2.0~\textnormal{g
cm}^{-2}$ exhibits low ejection efficiency due to during the initial
core-crossing of the outflow from $t=0$ to $t=0.3
t_\textsubscript{ff}$.  The ejection efficiency plateaus at later time
with $\epsilon_\textsubscript{core}$ varying between $0.34$ and $0.42$
at later time.  We note that by the end of the simulation at $t=0.8
t_{\textsubscript{ff}}$, the outflow ejection efficiency in the
simulation is 92\% of the analytic prediction.

We note that the star formation rate per free-fall time in our simulations,
$\epsilon_{\rm ff} = \epsilon_{\rm core} t_{\rm ff} / t_{\rm end}$, is much
greater than the observationally-estimated value of a few percent found by
\cite{KrumholzTan} and \cite{Evans}. This apparent discrepancy is
quite easy to understand, both observationally and theoretically.

On the observational side, the \cite{KrumholzTan} and \cite{Evans}
estimates were for gas clumps at densities of at most $\sim 10^5$ cm$^{-3}$,
and the typical objects at this density are $\sim 10^4$ $M_\odot$
parsec-sized clumps that are forming entire star clusters. In comparison,
the presetellar cores that we have simulated are much smaller and denser:
$n\sim 10^6$ cm$^{-3}$, $r \sim 0.1$ pc, $M \sim 10^2$ $M_\odot$. In the
terminology of \cite{McKeeOstriker}, they are ``cores'' rather than
``clumps''. There are no observational measurements for the value of
$\epsilon_{\rm ff}$ in such structures. Indeed, \cite{KrumholzTan}
commented that $\epsilon_{\rm ff}$ must reach values $\sim 1$ rather than
$\sim 0.01$ at some density higher than what then current observations probed.
Furthermore, it is clear that $100$ $M_\odot$ cores forming massive stars
must be rare exceptions, since massive stars are rare. Since measurements of
$\epsilon_{\rm ff}$ only provide statistical averages, it is possible that a
few $n \sim 10^6$ cm$^{-3}$ cores like the ones we have studied undergo
rapid collapse, but that there are more numerous structures at similar
density that are not undergoing rapid monolithic collapse, so that the
average value of $\epsilon_{\rm ff}$ is much lower than in the core we have
simulated.

On the theoretical side, values of $\epsilon_{\rm ff} \sim 0.01$ are
expected only in regions where there is turbulence at roughly virial levels
\citep{KrumholzOutflows}. In our simulations, while we start out with such
turbulence, this decays rapidly. Since these simulations contain a single
massive star with a dominant outflow, there is nothing to drive turbulence
in the core, and this allows $\epsilon_{\rm ff}$ to rise rapidly as the
turbulence becomes sub-virial. One can see this effect in figure 6: the
total mass in stars rises very slowly at first, and accelerates as time goes
on and the turbulence decays.

\section{Summary} \label{summary}
We report the results of several AMR radiation-hydrodynamic
simulations of the collapse of massive star forming clouds using the
ORION code.  These simulations are the first to include the feedback
effects of protostellar outflows, radiative heating, and radiation
pressure in a single computation.  In these simulations, the initial
density profile, velocity spectrum, virial ratio and numerical
resolution are held constant.  The simulations are scaled to different
surface density to study the environmental dependence of the outflow
and radiation feedback, and in one case the surface density is held
constant but outflow feedback is turned off to isolate the effect of
protostellar outflow.

Comparison of models with protostellar outflow feedback and surface
densities $\Sigma = 1.0~\textnormal{g cm}^{-2}$, $\Sigma = 2.0~\textnormal{g
cm}^{-2}$ and $\Sigma = 10.0~\textnormal{g cm}^{-2}$ at equivalent
free-fall time shows that the higher surface density clouds exhibit
enhanced radiative heating feedback, diminished disk fragmentation and
host more massive primary stars with less massive companions.
However, the effects of outflow feedback diminish with increased
surface density.  Lower surface density clouds have longer free-fall
time and therefore undergo more Kelvin contraction in the primary
protostellar core, leading to more powerful outflows and more
effective mechanical feedback.  Furthermore, lower surface density
clouds give rise to protostellar outflows with shorter core crossing
time relative to the core free-fall time and these clouds are
consequently influenced by the effect of outflows at relatively earlier
stages of collapse.

Comparison of models with and without outflow feedback at surface
density $\Sigma = 2.0~\textnormal{g cm}^{-2}$ indicates a strong coupling
between outflow and radiative feedback on the parent cloud.  Outflow
activity produces polar cavity of reduced optical depth through the
ambient core.  Radiation focusing in the direction of outflow cavities
is sufficient to prevent the formation of radiation pressure-supported
circumstellar gas bubbles, in contrast to models which neglect
protostellar outflow feedback.  With outflows, the radiative flux in
the poleward direction is enhanced by $1.7 \textnormal{ to } 15$ times
that in the midplane.  Sheets with outward radiative flux reduction up
to an order of magnitude appear near the equatorial latitude of the
primary star in all of the models with protostellar outflow.  This
result is consistent with the predictions of \cite{KrumholzOutflows}
that focusing of the radiative flux from the central star throughout
the outflow cavity results in a reduction of the equatorial radiative
flux relative to a control model without outflows by factors of $1.7$
to $14$, depending on the geometry of the outflow cavity.  As a
result the radiative heating and outward radiation force exerted on
the protostellar disk and infalling cloud gas in the equatorial
direction is greatly diminished by the presence of the outflow cavity,
and our models support the \cite{KrumholzOutflows} prediction that
outflows reduce the Eddington radiation pressure barrier to high-mass
star formation by reducing the radiation force exerted in the
infalling cloud gas.  Precisely determining the effect of outflows on the
threshold density prediction for massive star formation will require
examination of simulations at lower cloud surface density than the
models presented here.  Furthermore, we expect that the relative
importance of this effect may depend on the role of magnetic fields in
confining the outflow cavity.  Future works should therefore examine
the effect of protostellar outflows in lower surface density,
magnetized clouds.

\acknowledgements 
The authors are grateful for helpful discussions with
John Bally and the useful comments by the anonymous referee on the
topic of this paper.  Support for this work was provided by: the US
Department of Energy at the Lawrence Livermore National Laboratory
under contract DE-AC52-07NA 27344 (AJC and RIK); an Alfred P.\ Sloan
Fellowship (MRK); NASA through ATFP grant NNX09AK31G (RIK, CFM, and
MRK); NASA as a part of the Spitzer Theoretical Research Program,
through a contract issued by the JPL (MRK and CFM); the National
Science Foundation through grants AST-0807739 (MRK) and AST-0908553
(RIK and CFM). Support for computer simulations was provided by an
LRAC grant from the National Science Foundation through TeraGrid
resources, the Arctic Region Supercomputing Center (ARSC) and the NASA
Advanced Supercomputing Division.  The YT software toolkit
\citep{turk} was used for the data analysis and plotting.
LLNL-JRNL-472291.


\begin{thebibliography}{}
%
\bibitem[Arce et al.(2010)]{Arce} Arce, H.~G., Borkin, M.~A., Goodman, A.~A., Pineda, J.~E., \& Halle, M.~W.\ 2010, \apj, 715, 1170
%
\bibitem[Arce \& Sargent(2006)]{ArceSargent} Arce, H.~G., \& Sargent, A.~I.\ 2006, \apj, 646, 1070
%
%
\bibitem[Bachiller(1996)]{Bachiller} Bachiller, R.\ 1996, \araa, 34, 111 
%
\bibitem[Bertoldi \& McKee(1992)]{Bertoldi} Bertoldi, F., \& McKee, C.~F.\ 1992, \apj, 395, 140 
\bibitem[Beuther et al.(2007)]{Beuther07} Beuther, H., Leurini, S., Schilke, P., Wyrowski, F., Menten, K.~M., \& Zhang, Q.\ 2007, \aap, 466, 1065 
\bibitem[Beuther et al.(2004)]{Beuther04} Beuther, H., Schilke, P., \& Gueth, F.\ 2004, \apj, 608, 330 
\bibitem[Beuther et al.(2002a)]{Beuther02a} Beuther, H., Schilke, P., Gueth, F., McCaughrean, M., Andersen, M., Sridharan, T.~K., \& Menten, K.~M.\ 2002, \aap, 387, 931 
%
\bibitem[Beuther et al.(2002b)]{Beuther02b} Beuther, H., Schilke, P., Menten, K.~M., Motte, F., Sridharan, T.~K., \& Wyrowski, F.\ 2002, \apj, 566, 945 
\bibitem[Beuther et al.(2003)]{Beuther03} Beuther, H., Schilke, P., \& Stanke, T.\ 2003, \aap, 408, 601 
%
\bibitem[Beuther et al.(2002c)]{Beuther02c} Beuther, H., Walsh, A., Schilke, P., Sridharan, T.~K., Menten, K.~M., \& Wyrowski, F.\ 2002, \aap, 390, 289 
\bibitem[Beuther \& Shepherd(2005)]{Beuther05} Beuther, H., \& Shepherd, D.\ 2005, Cores to Clusters: Star Formation with Next Generation Telescopes, 105 
%
\bibitem[Carrasco-Gonz{\'a}lez et al.(2010)]{massivemhdjet} Carrasco-Gonz{\'a}lez, C., Rodr{\'{\i}}guez, L.~F., Anglada, G., Mart{\'{\i}}, J., Torrelles, J.~M., \& Osorio, M.\ 2010, Science, 330, 1209 
%
\bibitem[Carroll et al.(2010)]{Carroll} Carroll, J.~J., Frank, A., \& Blackman, E.~G.\ 2010, \apj, 722, 145 
%
\bibitem[Chabrier(2005)]{Chabrier} Chabrier, G.\ 2005, The Initial Mass Function 50 Years Later, 327, 41 
%
\bibitem[Chandler \& Richer(2000)]{Chandler} Chandler, C.~J., \& Richer, J.~S.\ 2000, \apj, 530, 851 
%
\bibitem[Crowther et al.(2010)]{Crowther} Crowther, P.~A., Schnurr, O., Hirschi, R., Yusof, N., Parker, R.~J., Goodwin, S.~P., \& Kassim, H.~A.\ 2010, \mnras, 408, 731 
%
\bibitem[Cunningham et al.(2006)]{Cunningham06} Cunningham, A.~J., Frank, A., \& Blackman, E.~G.\ 2006, \apj, 646, 1059
%
\bibitem[Curiel et al.(2006)]{Curiel} Curiel, S., et al.\ 2006, \apj, 638, 878 
%
\bibitem[Curtis et al.(2010)]{Curtis} Curtis, E.~I., Richer, J.~S., Swift, J.~J., \& Williams, J.~P.\ 2010, \mnras, 408, 1516 
%
\bibitem[Egan et al.(1998)]{egan} Egan, M.~P., Shipman, R.~F., Price, S.~D., Carey, S.~J., Clark, F.~O., \& Cohen, M.\ 1998, \apjl, 494, L199 
%
\bibitem[Evans et al.(2009)]{Evans} Evans, N.~J., II, et al.\ 2009, \apjs, 181, 321 
%
\bibitem[Ginsburg et al.(2011)]{Ginsburg} Ginsburg, A., Bally, J., \& Williams, J.\ 2011, arXiv:1106.1433 
\bibitem[Goodman et al.(1993)]{Goodman} Goodman, A.~A., Benson, P.~J., Fuller, G.~A., \& Myers, P.~C.\ 1993, \apj, 406, 528
%
\bibitem[Hatchell et al.(2007)]{Hatchell} Hatchell, J., Fuller, G.~A., Richer, J.~S., Harries, T.~J., \& Ladd, E.~F.\ 2007, \aap, 468, 1009 
%
\bibitem[Hennebelle et al.(2011)]{Hennebelle} Hennebelle, P., Commercon, B., Joos, M., Klessen, R.~S., Krumholz, M., Tan, J.~C., \& Teyssier, R.\ 2011, arXiv:1101.1574 
%
\bibitem[Hosokawa et al.(2011)]{Hosokawa} Hosokawa, T., Offner, S.~S.~R., \& Krumholz, M.~R.\ 2011, \apj, in press, arXiv:1101.3599 
%
\bibitem[Hutawarakorn et al.(2002)]{Hutawarakorn} Hutawarakorn, B., Cohen, R.~J., \& Brebner, G.~C.\ 2002, \mnras, 330, 349 
%
\bibitem[Jacquet \& Krumholz(2011)]{Jacquet} Jacquet, E., \& Krumholz, M.\ 2011, arXiv:1101.5265 
%
\bibitem[Jijina \& Adams(1996)]{Jijina} Jijina, J., \& Adams, F.~C.\ 1996, \apj, 462, 874 
%
\bibitem[Klein(1999)]{Klein} Klein, R.~I.\ 1999, Journal of Computational and Applied Mathematics, 109, 123 
%
\bibitem[Krumholz et al.(2010)]{krumholz10} Krumholz, M.~R., Cunningham, A.~J., Klein, R.~I., \& McKee, C.~F.\ 2010, \apj, 713, 1120 
%
\bibitem[Krumholz et al.(2007a)]{KrumholzStars} Krumholz, M.~R., Klein, R.~I., \& McKee, C.~F.\ 2007, \apj, 656, 959 
\bibitem[Krumholz et al.(2007b)]{KrumholzOrion} Krumholz, M.~R., Klein, R.~I., McKee, C.~F., \& Bolstad, J.\ 2007, \apj, 667, 626 
%
\bibitem[Krumholz et al.(2009)]{KrumholzRTBubbles} Krumholz, M.~R., Klein, R.~I., McKee, C.~F., Offner, S.~S.~R., \& Cunningham, A.~J.\ 2009, Science, 323, 754 
%
\bibitem[Krumholz \& McKee(2008)]{Krumholz08} Krumholz, M.~R., \& McKee, C.~F.\ 2008, \nat, 451, 1082 
%
\bibitem[Krumholz et al.(2005)a]{KrumholzOutflows} Krumholz, M.~R., McKee, C.~F., \& Klein, R.~I.\ 2005, \apjl, 618, L33 
%
\bibitem[Krumholz et al.(2004)]{KrumholzSinks} Krumholz, M.~R., McKee, C.~F., \& Klein, R.~I.\ 2004, \apj, 611, 399 
%
\bibitem[Krumholz \& Tan(2007)]{KrumholzTan} Krumholz, M.~R., \& Tan, J.~C.\ 2007, \apj, 654, 304 

%
\bibitem[Kuiper et al.(2010)]{Kuiper} Kuiper, R., Klahr, H., Beuther, H., \& Henning, T.\ 2010, \apj, 722, 1556 
%
\bibitem[Lada(2006)]{Lada} Lada, C.~J.\ 2006, \apjl, 640, L63 
%
\bibitem[Larson \& Starrfield(1971)]{larson} Larson, R.~B., \& Starrfield, S.\ 1971, \aap, 13, 190 
%
\bibitem[L{\'o}pez-Sepulcre et al.(2010)]{lopez} L{\'o}pez-Sepulcre, A., Cesaroni, R., \& Walmsley, C.~M.\ 2010, \aap, 517, A66 
%
\bibitem[L{\'o}pez-Sepulcre et al.(2011)]{lopez11} L{\'o}pez-Sepulcre, A., et al.\ 2011, \aap, 526, L2 
\bibitem[Matzner \& McKee(1999)]{MatznerOutflows} Matzner, C.~D., \& McKee, C.~F.\ 1999, \apjl, 526, L109
%
\bibitem[Matzner \& McKee(2000)]{MatznerCores} Matzner, C.~D., \& McKee, C.~F.\ 2000, \apj, 545, 364 
%
\bibitem[McCrady \& Graham(2007)]{McCrady} McCrady, N., \& Graham, J.~R.\ 2007, \apj, 663, 844 
%
\bibitem[McKee(1989)]{McKee89} McKee, C.~F.\ 1989, \apj, 345, 782 
%
\bibitem[McKee \& Ostriker(2007)]{McKeeOstriker} McKee, C.~F., \& Ostriker, E.~C.\ 2007, \araa, 45, 565 
\bibitem[McKee \& Tan(2002)]{McKeeTan02} McKee, C.~F., \& Tan, J.~C.\ 2002, \nat, 416, 59 
\bibitem[McKee \& Tan(2003)]{McKeeTan} McKee, C.~F., \& Tan, J.~C.\ 2003, \apj, 585, 850 
%
\bibitem[Maury et al.(2009)]{Maury} Maury, A.~J., Andr{\'e}, P., \& Li, Z.-Y.\ 2009, \aap, 499, 175 
%
\bibitem[Myers et al.(2011)]{myers} Myers, A.~T., Krumholz, M.~R., Klein, R.~I., McKee, C.~F., \& Tan, J.~C.\ 2011, in prep.
%
\bibitem[Nakamura \& Li(2007)]{Nakamura} Nakamura, F., \& Li, Z.-Y.\ 2007, \apj, 662, 395 
%
\bibitem[Nakano(1989)]{Nakano} Nakano, T.\ 1989, \apj, 345, 464 
%
\bibitem[Norman \& Silk(1980)]{NormanSilk} Norman, C., \& Silk, J.\ 1980, \apj, 238, 158
%
\bibitem[N{\"u}rnberger et al.(2007)]{Nurnberger} N{\"u}rnberger, D.~E.~A., Chini, R., Eisenhauer, F., Kissler-Patig, M., Modigliani, A., Siebenmorgen, R., Sterzik, M.~F., \& Szeifert, T.\ 2007, \aap, 465, 931 
%
\bibitem[Offner et al.(2009)]{Offner} Offner, S.~S.~R., Klein, R.~I., McKee, C.~F., \& Krumholz, M.~R.\ 2009, \apj, 703, 131 
%
\bibitem[Patel et al.(2005)]{Patel} Patel, N.~A., et al.\ 2005, \nat, 437, 109 
\bibitem[Pelletier \& Pudritz(1992)]{Pelletier92} Pelletier, G., \& Pudritz, R.~E.\ 1992, \apj, 394, 117 
%
\bibitem[Peters et al.(2010)]{Peters} Peters, T., Klessen, R.~S., Mac Low, M.-M., \& Banerjee, R.\ 2010, \apj, 725, 134
%
\bibitem[Perault et al.(1996)]{perault} Perault, M., et al.\ 1996, \aap, 315, L165 
%
\bibitem[Preibisch et al.(2001)]{Preibisch} Preibisch, T., Weigelt, G., \& Zinnecker, H.\ 2001, The Formation of Binary Stars, 200, 69 
%
\bibitem[Qiu et al.(2007)]{Qui} Qiu, K., Zhang, Q., Beuther, H., \& Yang, J.\ 2007, \apj, 654, 361 
%
\bibitem[Radhakrishnan \& Hindmarsh (1993)]{LSODE} Radhakrishnan, K. \&  Hindmarsh, A.~C.\ 1993, LLNL report UCRL-ID-113855.
%
\bibitem[Rathborne et al.(2006)]{rathborne} Rathborne, J.~M., Jackson, J.~M., \& Simon, R.\ 2006, \apj, 641, 389 
%
\bibitem[Rathborne et al.(2007)]{rathborne07} Rathborne, J.~M., Simon, R., \& Jackson, J.~M.\ 2007, \apj, 662, 1082 
\bibitem[Richer et al.(2000)]{Richer} Richer, J.~S., Shepherd, D.~S., Cabrit, S., Bachiller, R., \& Churchwell, E.\ 2000, Protostars and Planets IV, 867 
%
\bibitem[Rodriguez(1997)]{Rodriguez} Rodriguez, L.~F.\ 1997, Herbig-Haro Flows and the Birth of Stars, 182, 83 
%
\bibitem[Sandell \& Knee(2001)]{SandellKnee} Sandell, G., \& Knee, L.~B.~G.\ 2001, \apjl, 546, L49 
%
\bibitem[Saraceno et al.(1996)]{Saraceno} Saraceno, P., Andre, P., Ceccarelli, C., Griffin, M., \& Molinari, S.\ 1996, \aap, 309, 827 
%
\bibitem[Sch{\"o}nke \& Tscharnuter(2011)]{Schonke} Sch{\"o}nke, J., \& Tscharnuter, W.~M.\ 2011, \aap, 526, A139 
%
\bibitem[Semenov et al.(2003)]{Semenov} Semenov, D., Henning, T., Helling, C., Ilgner, M., \& Sedlmayr, E.\ 2003, \aap, 410, 611 
%
\bibitem[Shatsky \& Tokovinin(2002)]{Shatsky} Shatsky, N., \& Tokovinin, A.\ 2002, \aap, 382, 92 
%
\bibitem[Shepherd(2003)]{Shepherd} Shepherd, D.\ 2003, Galactic Star Formation Across the Stellar Mass Spectrum, 287, 333 
%
\bibitem[Shepherd \& Churchwell(1996)]{ShepherdChurchwell} Shepherd, D.~S., \& Churchwell, E.\ 1996, \apj, 472, 225 
%
\bibitem[Shu et al.(1987)]{shu87} Shu, F.~H., Adams, F.~C., \& Lizano, S.\ 1987, \araa, 25, 23 
\bibitem[Shu et al.(1988)]{Shu88} Shu, F.~H., Lizano, S., Ruden, S.~P., \& Najita, J.\ 1988, \apjl, 328, L19 
%
\bibitem[Swift(2009)]{Swift} Swift, J.~J.\ 2009, \apj, 705, 1456 
%
\bibitem[Toomre(1964)]{Toomre} Toomre, A.\ 1964, \apj, 139, 1217 
%
\bibitem[Torrelles et al.(1997)]{Torrelles} Torrelles, J.~M., Gomez, J.~F., Rodriguez, L.~F., Ho, P.~T.~P., Curiel, S., \& Vazquez, R.\ 1997, \apj, 489, 744 
%
\bibitem[Truelove(1997)]{Truelovethesis} Truelove, J.~K.\ 1997, Ph.D.~Thesis
%
\bibitem[Truelove et al.(1997)]{Truelove97} Truelove, J.~K., Klein, R.~I., McKee, C.~F., Holliman, J.~H., II, Howell, L.~H., \& Greenough, J.~A.\ 1997, \apjl, 489, L179 
%
\bibitem[Truelove et al.(1998)]{Truelove} Truelove, J.~K., Klein, R.~I., McKee, C.~F., Holliman, J.~H., II, Howell, L.~H., Greenough, J.~A., \& Woods, D.~T.\ 1998, \apj, 495, 821 
%
\bibitem[Turk et al.(2011)]{turk} Turk, M.~J., Smith, B.~D., Oishi, J.~S., Skory, S., Skillman, S.~W., Abel, T., \& Norman, M.~L.\ 2011, \apjs, 192, 9 
%
\bibitem[Turner et al.(2000)]{Turner} Turner, J.~L., Beck, S.~C., \& Ho, P.~T.~P.\ 2000, \apjl, 532, L109 
\bibitem[Walawender et al.(2005)]{Walawender05} Walawender, J., Bally, J., \& Reipurth, B.\ 2005, \aj, 129, 2308 
%
\bibitem[Wang et al.(2010)]{Wang} Wang, P., Li, Z.-Y., Abel, T., \& Nakamura, F.\ 2010, \apj, 709, 27 
%

\bibitem[Yorke \& Sonnhalter(2002)]{Yorke} Yorke, H.~W., \& Sonnhalter, C.\ 2002, \apj, 569, 846 
%
\bibitem[Zhang et al.(2001)]{Zhang} Zhang, Q., Hunter, T.~R., Brand, J., Sridharan, T.~K., Molinari, S., Kramer, M.~A., \& Cesaroni, R.\ 2001, \apjl, 552, L167 
%
\bibitem[Zinnecker \& Yorke(2007)]{Zinnecker} Zinnecker, H., \& Yorke, H.~W.\ 2007, \araa, 45, 481 
%
\end{thebibliography}
\end{document}